\documentclass[preprint,12pt,authoryear,review]{elsarticle}
\usepackage[a4paper]{geometry}
\usepackage[titletoc]{appendix}

\usepackage{tikz}

\usepackage{amssymb}
\usepackage{amsfonts}
\usepackage{mathrsfs} % Ralph Smith’s Formal Script Font (rsfs) => with command \mathscr{}
\usepackage{amsmath, amsthm, amssymb}
\usepackage{graphicx}
\usepackage{hyperref, comment}
\usepackage{bbm}
\usepackage{pgf}

\usetikzlibrary{fit}					% fitting shapes to coordinates
\usetikzlibrary{backgrounds}	
\newtheorem{theorem}{Theorem}%[section]
\newtheorem{lemma}[theorem]{Lemma}

\newtheorem{proposition}[theorem]{Proposition}

\newtheorem{corollary}[theorem]{Corollary}
\newtheorem{definition}[theorem]{Definition}
\newtheorem{example}[theorem]{Example}
\newtheorem{discussion}[theorem]{Discussion}
\newtheorem{remark}[theorem]{Remark}

\definecolor{darkblack}{rgb}{0, .07, .5}
\definecolor{darkred}{rgb}{0.5,0,0}
 \definecolor{mahogany}{rgb}{0.65, 0., 0.5}

\def\Pr{\text{\rm{Pr}}}

\newcommand{\bR}{\mathbb{R}}

\newcommand{\E}{\mathbb{E}}

\newcommand{\RN}[1]{%
  \textup{\uppercase\expandafter{\romannumeral#1}}%
}
\journal{Games and Economic Behavior}

\begin{document}

\begin{frontmatter}


\newcommand\blfootnote[1]{%
  \begingroup
  \renewcommand\thefootnote{}\footnote{#1}%
  \addtocounter{footnote}{-1}%
  \endgroup
}
\title{Playing Games with Bounded Entropy: Convergence Rate and Approximate Equilibria}
%\title{Playing Games with Bounded Entropy: Convergence Rate and Approximate Equilibria\tnoteref{t1}}
%\tnotetext[t1]{This work was supported by Sharif University of Technology under Grant QB950607.}


\author[1]{Mehrdad Valizadeh}
\ead{valizadeh@ee.sharif.edu}

\author[1]{Amin Gohari\corref{cor}}
\ead{aminzadeh@sharif.edu}
\cortext[cor]{Corresponding author}

\address[1]{Department of Electrical Engineering, Sharif University of Technology, Tehran, Iran}


\begin{abstract}
We consider zero-sum repeated games in which the players are restricted to strategies that require only a limited amount of randomness. Let $v_n$ be the max-min value of the $n$ stage game; previous works have characterized $\lim_{n\rightarrow\infty}v_n$, \emph{i.e.,} the long-run max-min value. Our first contribution is to study the convergence rate of $v_n$ to its limit. To this end, we provide a new tool for simulation of a source (target source) from another source (coin source). Considering the total variation distance as the measure of precision, this tool offers an upper bound for the precision of simulation, which is vanishing exponentially in the difference of R\'enyi entropies of the coin and target sources. In the second part of paper, we characterize the set of all approximate Nash equilibria achieved in long run. It turns out that this set is in close relation with the long-run max-min value.
\end{abstract}

\begin{keyword}


Repeated Games \sep Bounded Entropy \sep Randomness Extraction \sep Source Simulation \sep Information Theory
\end{keyword}

\end{frontmatter}



\tableofcontents
\section{Introduction}\label{S:intro}
\cite{Nash} showed that all one-shot games have at least one equilibrium in the mixed strategies. Private randomness is required to implement mixed strategies, and consequently
a Nash equilibrium may not exist if insufficient random bits are available to the players (See \cite{Hubacek} and \cite{Budinich}).

Limited randomness in repeated zero-sum games was originally studied by \cite{Neyman2000} and \cite{Gossner2002}. \cite{Gossner2002} studied a repeated zero-sum game between Alice (the maximizer) and Bob (the minimizer). At the beginning of each stage of the game, Alice observed an independent drawing of a random source $X$ with a commonly known distribution. Next, the players played an action which was monitored by the other player. The only source of randomization available to Alice was the outcomes of random source $X$. Thus, Alice had to choose the action of each stage as a \emph{deterministic} function of the history of her observations, \emph{i.e.,} the random sources revealed up to that stage and the previous actions. However, Bob could freely randomize his actions, and hence, at each stage, he chose his action as a \emph{random} function of the actions played previously. Generalizing the model of \cite{Gossner2002}, \cite{bounded_entropy} considered the possibility of leakage of Alice's random source sequence to Bob; thus, they called it the \emph{repeated game with leaked randomness source}. In other words, Bob monitored the random source of Alice through a noisy channel. Specifically, let $(X_1, Y_1)$, $(X_2, Y_2), \ldots$ be a sequence of independent and identically distributed (\emph{i.i.d.\@}) random variables distributed according to a given distribution $p_{XY}$. At arbitrary stage $t$, before choosing the actions for that stage, Alice observed $X_t$, and Bob observed $Y_t$. In this model, Alice and Bob could randomize their actions at each stage just by conditioning their actions to the history of their observations up to that stage.

In this paper, we study two different aspects of the repeated game with leaked randomness sources. Our first contribution is to study the max-min payoff that Alice can secure in a repeated game with finite number of stages. Note that \cite{bounded_entropy} characterized the long run max-min value, \emph{i.e.,} the maximum payoff that Alice can secure regardless of what strategy Bob chooses when the number of stages tends to infinity. More precisely, let $v_n$ be the max-min value of the $n$-stage repeated game with leaked randomness source. \cite{bounded_entropy} characterized $\lim_{n\to \infty} v_n$. In this paper, we investigate how $v_n$ converges to its limit. To do so, we develop and utilize a new tool for simulation of a source from another source, which we will introduce later in Section~\ref{sec:intro_simulation}.

Our second contribution is to study the set of equilibria that is implementable by Alice and Bob in the repeated game with leaked randomness sources. As stated above, implementable Nash equilibria do not necessarily exist. However, a relaxed version of Nash equilibria called approximate Nash equilibria may exist. Let $\epsilon_A$ and $\epsilon_B$ be arbitrary positive numbers. We say a given strategy profile forms a $(\epsilon_A,\epsilon_B)$-Nash equilibrium if Alice and Bob do not gain more than $\epsilon_A$ and $\epsilon_B$, respectively, by unilaterally changing their corresponding strategies. We characterize the set of $(\epsilon_A,\epsilon_B)$-Nash equilibria of the repeated game when the number of stages of the game tends to infinity. This set is characterized in terms of the maximum payoffs that Alice and Bob can secure in long run (long run max-min and min-max values).

Note that in previous works (\cite{Neyman2000, Gossner2002, bounded_entropy}), the max-min (or min-max) value of the zero-sum repeated game was achieved by autonomous strategies -- a strategy that is indifferent about the actions of the opponent in past stages. Therefore, we address the question as to whether autonomous strategies are sufficient for achieving all implementable approximate Nash equilibria. To do this, we also characterize the set of all approximate Nash equilibria achieved by autonomous strategies in long run. It will turn out that the set of approximate equilibria achieved by autonomous strategies is absolutely smaller than the set of approximate equilibria achieved by arbitrary strategies.

\subsection{A new tool}\label{sec:intro_simulation}
A key step in the proofs of \cite{Gossner2002} and \cite{bounded_entropy} is to divide the total $n$ stages of the repeated game into some blocks such that the actions of the first player in each block (excluding the first block) is generated as a function of the randomness source observed during the previous block.\footnote{This strategy is known as the \emph{block Markov} strategy in information theory and utilized in multi-hop communication settings.} In other words, the actions of the first player in each block is \emph{simulated} from the randomness source observed in the previous block. Since we are interested in the non-asymptotic regime where the number of stages $n$ is given and fixed, we need to carefully optimize over the length of the blocks and also prove a fine estimate on the accuracy of simulation of the actions of each block from the observations of the previous block. Thus, in order to study the repeated game with $n$ stages, we provide a new tool for simulation of a source from another source which is of independent interest. 

More precisely, in abstract terms, let $X\in\mathcal X$ and $Y\in\mathcal Y$ be arbitrary discrete random variables distributed according to some probability mass function $p_{XY}$, and let $A\in \mathcal A$ be a target random variable distributed according to $p_A$. We would like to simulate $A$ from $X$ (by using a deterministic function $f:\mathcal X\to \mathcal A$) in such a way that the resulting random variable, $f(X)$, is almost independent of $Y$, and its distribution is close to $p_A$. Intuitively, if the amount of uncertainty of $X$ given $Y$ is much more than the amount of uncertainty of $A$, then, one might find a simulator $f:\mathcal X\to \mathcal A$ satisfying the above conditions. We take the R\'enyi entropy as our measure of uncertainty, and the total variation distance as our measure of similarity. We prove that there exists a mapping $f:\mathcal X\to \mathcal A$ such that for arbitrary $1\leq \alpha\leq 2$,
\begin{equation}\|p_{f(X)Y}-p_Ap_Y\|_{TV} \leq 2^{-(1-\frac{1}{\alpha})\left(H_{\alpha}(X|Y)-H_{\frac{1}{\alpha}}(A)+2\right)},\label{eqnNew1}\end{equation}
where $\|.\|_{TV}$ denotes the total variation distance, $H_{\alpha}(X|Y)$ denotes the conditional R\'enyi entropy (with parameter $\alpha$) of $X$ given $Y$, and $H_{\frac{1}{\alpha}}(A)$ is the R\'enyi entropy of $A$ with parameter $1/\alpha$. The main idea to prove Equation \eqref{eqnNew1} is to relate it to norms of linear maps and then utilize the Riesz-Thorin Interpolation Theorem.

To better understand Equation \eqref{eqnNew1}, let us apply it to a sequence of random variables.
Assume that $(X_1, Y_1)$, $(X_2, Y_2)$, \ldots, $(X_n, Y_n)$ are $n$ \emph{i.i.d.\@} repetitions according to $p_{XY}$. Our goal is to simulate $(A_1, A_2,\ldots, A_n)$, which is an \emph{i.i.d.\@} sequence according to $p_A$. Applying Equation~\eqref{eqnNew1} to $\tilde{X}=(X_1, \ldots, X_n)$, $\tilde{Y}=(Y_1, \ldots, Y_n)$, and $\tilde{A}=(A_1,\ldots, A_n)$, we obtain that there exists a mapping $f:\mathcal X^n\to \mathcal A^n$ such that for arbitrary $1\leq \alpha\leq 2$,
\begin{align}\|p_{f(X^n)Y^n}-p_{A^n}p_{Y^n}\|_{TV} &\leq 2^{-(1-\frac{1}{\alpha})\left(nH_{\alpha}(X|Y)-nH_{\frac{1}{\alpha}}(A)+2\right)}\nonumber
\\&\leq  2^{-n(1-\frac{1}{\alpha})\left(H_{\alpha}(X|Y)-H_{\frac{1}{\alpha}}(A)\right)},\label{eqnNew22}\end{align}
where we used the fact that $H_{\alpha}(X^n|Y^n)=n H_{\alpha}(X|Y)$ and $H_{\frac{1}{\alpha}}(A^n)=nH_{\frac{1}{\alpha}}(A)$. Equation~\eqref{eqnNew22} shows that the accuracy of simulation is improving exponentially fast in the product of three terms: the block length $n$, the term $1-{1}/{\alpha}$, and the entropy difference $H_{\alpha}(X|Y)-H_{\frac{1}{\alpha}}(A)$.

Moreover, Equation~\eqref{eqnNew1} can be interpreted in a different way: we say that $\mathsf{R}(\cdot)$ is a \emph{measure of randomness} if for any discrete random variable $X$, $\mathsf{R}(X)$ is a non-negative real number. The value $\mathsf{R}(X)$ quantifies the amount of uncertainty in $X$. Then, $\mathsf{R}(\cdot)$ is a reasonable measure of randomness only if it is non-increasing under mappings. In other words, if random variable $A$ is a deterministic function of random variable $X$, we expect $\mathsf{R}(X)\geq \mathsf{R}(A)$. The question then arises whether the converse to this statement can also be true:

\noindent\underline{Question:} \emph{Is there a suitable measure of randomness $\mathsf{R}(\cdot)$ such that $\mathsf{R}(X)\geq \mathsf{R}(A)$ \emph{if and only if} there is a function $f:\mathcal{X}\mapsto\mathcal{A}$ such that $f(X)$ is distributed according to $p_A$?}

\noindent While the answer to this question is negative, our tool shows that an approximate version of it holds. To see why the answer to this question is negative, let $X\in\{0,1\}$ be a binary random variable. Then, $f(X)$ has the same amount of randomness as $X$ if $f$ is a one-to-one function ($f(0)\neq f(1)$), and $f(X)$ is deterministic if $f(0)=f(1)$. Therefore, $\mathsf{R}(f(X))\in \{0, \mathsf{R}(X)\}$ and cannot take values lying between $0$ and $\mathsf{R}(X)$. However, if we require $f(X)$ to have a distribution that is ``approximately" equal to $p_A$, the above question can be revisited. In fact, our tool shows that R\'enyi entropy is an answer for the approximate version of the above question.

\noindent
\textbf{Relation of Equation \eqref{eqnNew1} to previous works:} The problem of simulation of a source from another source dates back to the work of \cite{von_neumann_1951}, who considered the problem of generating a sequence of \emph{i.i.d.\@} fair bits from a given sequence of \emph{i.i.d.\@} unfair bits. The algorithm presented by \cite{von_neumann_1951} is \emph{universal} in the sense that it does not need the knowledge of the distribution of the input bits, and it is \emph{exact} in the sense that the output bits are exactly fair. \cite{von_neumann_1951} also offered a non-universal exact algorithm for simulation of a desired continuous distribution from a given continuous random variable with known distribution. A generalization of the algorithm of \cite{von_neumann_1951} for arbitrary Markov inputs can be found in \cite{elias_1972} and \cite{generalized_elias}. There are other works that have considered non-exact simulation of a source. Considering the total variation distance as the measure of accuracy, \cite{yasayi} studied non-universal generation of independent fair bits from an \emph{i.i.d.\@} sequence of random variables with side information, and \cite{han} considered the simulation of a general sequence from a general input sequence with known distribution. Fundamental limits for generation of arbitrary random sequence from a general sequence of random variables under different measures of accuracy has been studied by \cite{verdu_1995} and \cite{yu_2019}.

Above works considered the simulation of an intended long sequence from a long input sequence. In contrast, a different approach for generating random bits (randomness extraction) is to provide results for arbitrary single-letter sources, and then, conclude results for sequences; works of \cite{renner}, \cite{hayashi2011exponential} and \cite{mojahedian2018correlation} on randomness extraction and privacy amplification lie in this category. The tool we present in this paper generalizes the results of \cite{renner}, \cite{hayashi2011exponential} and \cite{mojahedian2018correlation}; in fact, they considered the special case of simulation of random variable $A$ having a \emph{uniform distribution} over a set $\mathcal A$ (when $A$ is uniform, simulating $A$ can be interpreted as extracting $\log |\mathcal{A}|$ bits of randomness). Furthermore, in this paper, we adopt the total variation distance as the measure of accuracy which has a close relation with the expected payoff in games. We also use concentration inequalities to provide further refinements (Proposition \ref{pro:simulation}).

The rest of this paper is organized as follows: In Section~\ref{sec:pre}, we introduce the notations of this paper and present a brief discussion of Shannon and R\'enyi entropy. The repeated game with leaked randomness source is defined in Section~\ref{sec:leaked1}, where we also provide our results on the convergence rate of the max-min payoff of games with finite number of stages. In Section~\ref{sec:simulation}, we introduce our tool for simulation of a source from another source. In Section~\ref{sec:leaked2}, we characterize the set of approximate Nash equilibria achievable in long run. Some of the proofs are presented in Appendices.

\section{Preliminaries}\label{sec:pre}
\subsection{Notations} \label{def-sec-nkl4}
In this paper, we use the notation $x^j$ to represent a sequence of variables $(x_1,x_2,\ldots,x_j)$. The same notation is used to represent sequences of random variables, \emph{i.e.,} $X^j=(X_1,X_2,\ldots,X_j)$. Note that this notation is used for sequences that have two subscripts the same way, \emph{i.e.,} $X_k^j=(X_{k,1},X_{k,2},\ldots,X_{k,j})$. Calligraphic letters such as $\mathcal{X},\mathcal{Y}, \mathcal{A}, \mathcal{B}, \dots$ represent finite sets, and $|\mathcal X|$ denotes the cardinality of the finite set $\mathcal X$. Cartesian product of two sets $\mathcal A$ and $\mathcal B$ is denoted by $\mathcal A\times \mathcal B$, and $\mathcal A^n$ stands for $n$ times cartesian product of $\mathcal A$. The set of natural numbers is represented by $\mathbbm N$, and $\mathbbm R$ denotes the set of real numbers. For a real number $a$, $\lfloor a \rfloor$ is the largest integer less than or equal to $a$, and $\lceil a \rceil$ is the smallest integer greater than or equal to $a$. Furthermore, let $f(.)$ and $g(.)$ be two real functions on the set of real numbers; we write $f(a)=\mathcal O(g(a))$ if and only if there exists a real constant $c$ such that for all $a\in \mathbbm R$, we have $|f(a)|\leq c|g(a)|$. We use the notation $f_n=\mathcal O(g_n)$ for real sequences $\{f_n\}_{n\in \mathbbm N}$ and $\{g_n\}_{n\in \mathbbm N}$ in the same manner.

The probability mass function (pmf) of a random variable $X$ is represented by $p_X(x)$. When it is obvious from the context, we drop the subscript and use $p(x)$ instead of $p_X(x)$. We say that $X^n$ is drawn \emph{i.i.d.\@} from $p(x)$ if
$$p(x^n)=\prod_{i=1}^np(x_i).$$
We use $\Delta (\mathcal{A})$ to denote the probability simplex on alphabet $\mathcal{A}$, \emph{i.e.,} the set of all probability distributions on the finite set $\mathcal{A}$.
The total variation distance between pmfs $p_X$ and $q_X$ is denoted by $\|p_X-q_X\|_{TV}$ and is defined as:
$$\|p_X-q_X\|_{TV}\triangleq \frac 12 \sum_{x\in \mathcal X}|p_X(x)-q_X(x)|.$$
Some of the properties of the total variation distance are summarized in the following lemma.
\begin{lemma}\label{lemma:tv_p}
The following properties hold for the total variation distance:
\begin{description}
\item[\quad\textbf{Property 1:}] $\|p_{E}p_{F|E}-q_{E}p_{F|E}\|_{TV}=\|p_{E}-q_{E}\|_{TV}$;
\item[\quad\textbf{Property 2:}] $\|p_{E}p_{F|E}-q_{E}q_{F|E}\|_{TV}\geq \|p_{E}-q_{E}\|_{TV}$;
\item[\quad\textbf{Property 3:}] $\|p_{E_1}p_{F_1}-p_{E_2}q_{F_2}\|_{TV}\leq \|p_{E_1}-p_{E_2}\|_{TV}+\|p_{F_1}-q_{F_2}\|_{TV}$.
\end{description}
\end{lemma}
\subsection{Shannon Entropy}

Let $X\in \mathcal{X}$ and $Y\in \mathcal Y$ be two random variables with joint probability distribution $p_{XY}$ and respective marginal distributions $p_X$ and $p_Y$. The Shannon entropy of the random variable $X$ is defined to be:
$$H(X)=\sum_{x\in \mathcal X} -p_X(x) \log(p_X(x)),$$
where $0\log(0)=0$ by continuity, and all logarithms in this paper are in base two. Since the Shannon entropy is a function of the pmf $p_X$, we sometimes write $H(p_X)$ instead of $H(X)$.

The conditional Shannon entropy of $X$ given $Y$ is defined as:
\begin{align*}
H(X|Y)&=\sum_{(x,y)\in\mathcal{X}\times \mathcal{Y}} -p_{XY}(x,y) \log(p_{X|Y}(x|y))\\
&=\sum_{y\in \mathcal Y} p_Y(y)H(X|Y=y),
\end{align*}
where $H(X|Y=y)=\sum_{x\in \mathcal X} -p_{X|Y}(x|y) \log(p_{X|Y}(x|y))$.

The following properties hold for the entropy function:
\begin{itemize}
\item $H(X)\geq 0.$
\item For arbitrary deterministic function $f(x)$, we have $H(f(X))\leq H(X)$.
\end{itemize}

\subsection{R\'enyi Entropy}

Let $X\in \mathcal{X}$ and $Y\in \mathcal Y$ be two random variables with joint probability distribution $p_{XY}$ and respective marginal distributions $p_X$ and $p_Y$. For arbitrary $\alpha>0$, the R\'enyi entropy of random variable $X$ with parameter $\alpha$ is defined as follows:
$$H_{\alpha}(X)=\frac{\alpha}{1-\alpha}\log\left(\left(\sum_{x\in \mathcal X} p_X(x)^{\alpha}\right)^{\frac{1}{\alpha}}\right) = \frac{\alpha}{1-\alpha}\log\|p_X\|_{\alpha},$$
where $\|p_X\|_{\alpha}=\left(\sum_{x\in \mathcal X} p_X(x)^{\alpha}\right)^{\frac{1}{\alpha}}$ is the $\alpha$-norm of $p_X$.
Since the R\'enyi entropy is a function of the pmf $p_X$, we sometimes write $H_{\alpha}(p_X)$ instead of $H_{\alpha}(X)$.

The conditional R\'enyi entropy of $X$ given $Y$ with parameter $\alpha$ is defined as:
$$H_{\alpha}(X|Y)=\frac{\alpha}{1-\alpha}\log\left(\sum_{y\in \mathcal Y}p_Y(y)\|p_{X|Y=y}\|_{\alpha}\right),$$
where $p_{X|Y=y}$ is the conditional distribution of $X$ given $Y=y$.

R\'enyi entropy is related to Shannon entropy by the following relations:
$$\lim_{\alpha\to 1} H_{\alpha}(X) = H(X),\quad \lim_{\alpha\to 1} H_{\alpha}(X|Y)=H(X|Y).$$

Let us fix $X\in \mathcal X$ and consider $H_{\alpha}(X)$ as a function of $\alpha$. $H_{\alpha}(X)$ is analytic for all $\alpha\geq 0$, and hence, differentiable of all orders. In this paper, we are interested in the values of R\'enyi entropy for $1/2\leq \alpha\leq 2$. Particularly, for $\alpha=1$ we have:
$$d_1(X)\triangleq -\frac{d}{d_{\alpha}}H_{\alpha}(X)\Big |_{\alpha=1}=\frac{1}{2\log e} \left(\sum_{x\in \mathcal X}p(x)(\log (p(x)))^2-H(X)^2\right).$$
Note that $H(X)=\sum_{x\in \mathcal X}p(x)\log (p(x))$, and function $f(a)=a^2$ is convex. Therefore, Jensen's inequality implies that $d_1(X)$ is non-negative. Using the Taylor expansion, for $1/2\leq \alpha\leq 2$, we have:
\begin{align}H_{\alpha}(X)=H(X)-d_1(X)(\alpha-1)+ R_X(\alpha),\label{eqnO1}\end{align}
where the remainder $R_X(\alpha)$ is bounded as 
\begin{align}
|R_X(\alpha)|\leq d_2(X)(\alpha-1)^2,
\end{align}\label{eqnO2}
where
$$d_2(X)=\frac 12\textrm{   } \underset{1/2\leq \alpha' \leq 2}{\max\textrm{ }}\left|\frac{d^2 H_{\alpha}(X)}{d\alpha^2}\Big|_{\alpha=\alpha'}\right|.$$
Since $d_1(X)$ and $d_2(X)$ are functions of $p_X$, instead of them, we will sometimes write $d_1(p_X)$ and $d_2(p_X)$, respectively. Similarly, for the conditional R\'enyi entropy and for $1/2\leq \alpha\leq 2$, we have
\begin{align}
H_{\alpha}(X|Y)=H(X|Y)-d_1(X|Y)(\alpha-1)+R_{X|Y}(\alpha),\label{eqnO3}
\end{align}
where $R_{X|Y}(\alpha)$ is the remainder term, and 
\begin{align}
d_1(X|Y)= -\frac{d}{d_{\alpha}}H_{\alpha}(X|Y)\Big |_{\alpha=1}=\sum_{y\in \mathcal Y}p_Y(y)d(p_{X|Y=y}) + \frac{1}{2\log e} \left(\sum_{y\in \mathcal Y}p_Y(y)H(X|Y=y)^2-H(X|Y)^2\right).\notag
\end{align}
Again, Jensen's inequality implies that $d_1(X|Y)$ is non-negative. Moreover, the remainder $R_{X|Y}(\alpha)$ is bounded as 
\begin{align}
|R_X(\alpha)|\leq d_2(X|Y)(\alpha-1)^2,\label{eqnO4}
\end{align}
where
$$d_2(X|Y)=\frac 12 \underset{1/2\leq \alpha' \leq 2}{\max}\left|\frac{d^2 H_{\alpha}(X|Y)}{d\alpha^2}\Big|_{\alpha=\alpha'}\right|.$$

A more detailed analysis of the R\'enyi entropy with respect to the parameter $\alpha$ can be found in \cite [Section 5]{beck1995thermodynamics}.

\section{Repeated games with leaked randomness source: convergence rate}\label{sec:leaked1}
In this section, we revisit the repeated game of \cite{Gossner2002}. Here, we focus on its general version with a leaked randomness source studied by \cite{bounded_entropy}. \cite{bounded_entropy} characterized the max-min value of the repeated game when the number of the stages of the game tends to infinity. In contrast, we let the number of stages of the game be fixed to $n\in \mathbbm N$, and investigate the rate by which the max-min value of the $n$-stage game converges to the long-run max-min value.

\subsection{Problem statement and results}\label{sec:bounded_entropy}
Consider an $n$ stage repeated zero-sum game between players Alice($A$) and Bob($B$) with respective pure action sets $\mathcal{A}$ and $\mathcal{B}$. Let $\mathcal X$ and $\mathcal Y$ be the alphabet of randomness sources of Alice and Bob, respectively, and let $p_{XY}$ be a publicly known pmf on $\mathcal X\times \mathcal Y$. At each stage $t\in\{1,2,\dots,n\}$, random variables $X_t\in \mathcal X$ and $Y_t\in \mathcal Y$ are drawn independent of previous drawings according to $p_{XY}$, where $X_t$ is observed by Alice and $Y_t$ is observed by Bob. Then, Alice and Bob choose respective actions $A_t\in \mathcal{A}$ and $B_t\in\mathcal{B}$. At the end of stage $t$, players monitor the chosen actions $A_t$ and $B_t$, and Alice gets stage payoff $u_{A_t B_t}$ from Bob. In order to choose $A_t$ and $B_t$, players use the history of their observations until stage $t$. Let $\mathsf H_1^t=(X_{1},A_1,B_1, \dots,X_{t-1},A_{t-1},B_{t-1},X_{t})$ and $\mathsf H_2^t=(Y_{1},A_1,B_1,\dots,Y_{t-1},A_{t-1},B_{t-1},Y_{t})$ denote the history of observation of Alice and Bob (respectively) up to stage $t$. Then, $A_t=\sigma_t(\mathsf H_1^t)$ and $B_t=\tau_t(\mathsf H_2^t)$, where $\sigma_t:(\mathcal{A}\times\mathcal{B})^{t-1}\times  \mathcal{X}^t\to \mathcal{A}$ and $\tau_t:(\mathcal{A}\times\mathcal{B})^{t-1}\times  \mathcal{Y}^t\to \mathcal{B}$ are deterministic functions
by which Alice and Bob map their observations into their actions at stage $t$. Notice that the mappings $\sigma_t$ and $\tau_t$ are deterministic which means that the only source of randomization are $\mathsf H_1^t$ (for Alice) and $\mathsf H_2^t$ (for Bob). We call the $n$-tuples $\sigma^n=(\sigma_1,\sigma_2,\dots,\sigma_n)$ and $\tau^n=(\tau_1,\tau_2,\dots,\tau_n)$ the strategies of Alice and Bob, respectively. The expected average payoff for Alice up to stage $n$ induced by strategies $\sigma^n$ and $\tau^n$ is denoted by $\lambda(\sigma^n,\tau^n)$:
\begin{equation}\label{eq:exp_avg}
\lambda(\sigma^n,\tau^n)=\E_{\sigma^n,\tau^n}\left[\frac{1}{n}\sum_{t=1}^n u_{A_tB_t}\right],
\end{equation}
where $\E_{\sigma^n,\tau^n}$ denotes the expectation with respect to the distribution induced by \emph{i.i.d.\@} repetitions of $p_{XY}$ and strategies $\sigma^n$ and $\tau^n$. Alice wishes to maximize $\lambda(\sigma^n,\tau^n)$ and Bob's goal is to minimize it.

We will refer to the above game with \emph{``the repeated game with leaked randomness source''}. Another variant of this game, called \emph{``the repeated game with non-causal leaked randomness source''} is defined in the following remark.
\begin{remark}\label{remark2}
In the definition of the repeated game with leaked randomness source, we assumed that the randomness sources $X^n=(X_1,\dots,X_n)$ and $Y^n=(Y_1,\dots,Y_n)$ are revealed to Alice and Bob causally as the game is played out. However, we can also consider the non-causal case in which the sources $X^n$ and $Y^n$ are observed by Alice and Bob (respectively) before the game starts. In this case we have $\mathsf H_1^t=(X^n,A^{t-1},B^{t-1})$ and $\mathsf H_2^t=(Y^n,A^{t-1},B^{t-1})$. In order to distinguish the above two cases, we name the non-causal game as ``the repeated game with non-causal leaked randomness source''.
\end{remark}

\begin{definition}\label{def:max-min value}
Let $v$ be an arbitrary real value:
\begin{itemize}
\item Alice can secure $v$ in the $n$ stage repeated game if there exists a strategy $\sigma^n$ for Alice such that for all strategy $\tau^n$ of Bob we have $\lambda(\sigma^n,\tau^n) \geq v$. The maximum of the set of payoffs $v$ that Alice can secure in the $n$ stage repeated game is called the max-min value of the $n$-stage game.
\item Alice can secure $v$ in long run if there exists a sequence of strategies $\{\sigma^n\}_{n\in \mathbbm N}$ for Alice such that for all sequences of strategies $\{\tau^n\}_{n\in \mathbbm N}$ of Bob we have $\liminf_{n\to \infty} \lambda(\sigma^n,\tau^n) \geq v$. The supremum of the set of payoffs $v$ that Alice can secure in long run is called the long run max-min value of the game.
\end{itemize}
\end{definition}

The set of all payoffs that can be secured in long run in the repeated game with leaked randomness source is characterized by \cite{bounded_entropy} and restated here as Theorem~\ref{theorem:maxmin_val}. Before presenting Theorem~\ref{theorem:maxmin_val}, we need the following definition.
\begin{definition}\label{def:security_level}
In a stage game, the security level of mixed action $p_A$ for Alice is denoted by $U^{(A)}(p_A)$, and is defined as follows:
\begin{equation}\label{eq:u}
U^{(A)}(p_A)=\min_{b\in\mathcal{B}} \sum_{a\in \mathcal A}p_A(a)u_{ab}.
\end{equation}
Furthermore, the maximum payoff that Alice can secure in a stage game, by playing mixed actions of entropy at most $\mathsf h$, is denoted by $\mathcal{J}^{(A)}(\mathsf h)$, and is defined as:
\begin{equation}\label{eq:j}
\mathcal{J}^{(A)}(\mathsf h) = \max_{p_A\in \Delta(\mathcal{A}), H(p_A)\leq \mathsf{h}} U^{(A)}(p_A).
\end{equation}
\end{definition}

\begin{theorem}[\cite{bounded_entropy}]\label{theorem:maxmin_val}
Let $\mathcal J^{(A)}_{\text{cav}} (\mathsf h)$ be the upper concave envelope of $\mathcal{J}^{(A)}(\mathsf h)$ defined in Definition~\ref{def:security_level}. In the repeated game with leaked randomness source, Alice can secure $v$ in long run if and only if $v\leq\mathcal J^{(A)}_{\text{cav}} (H(X|Y))$. Furthermore, in $n\in \mathbbm{N}$ stage game, Alice can secure $v$ only if $v\leq \mathcal J^{(A)}_{\text{cav}} (H(X|Y))$.
\end{theorem}

Theorem~\ref{theorem:maxmin_val} implies that the long run max-min value of the repeated game with leaked randomness source is $\mathcal J^{(A)}_{\text{cav}} (H(X|Y))$. Moreover, the max-min value of the $n$-stage game is at most $\mathcal J^{(A)}_{\text{cav}} (H(X|Y))$. In the following theorems we discuss how the max-min value of the $n$-stage game converges to $\mathcal J^{(A)}_{\text{cav}} (H(X|Y))$ as $n$ increases.

\begin{theorem}\label{theorem:finite}
In the repeated game with leaked randomness source, there exist real numbers $r>0$, $\beta>0$, $\gamma\geq0$ and $\mu\geq 0$, such that the following property holds: for arbitrary sequences $\{f_n\}_{n\in \mathbbm N}$, $\{g_n\}_{n\in \mathbbm N}$ and $\{h_n\}_{n\in \mathbbm N}$ satisfying
$f_n\in \mathbbm N$, $0 \leq g_n\leq r$, and $0\leq h_n\leq 1$,
one can find a sequence of strategies $\{\sigma^n\}_{n\in \mathbbm N}$ such that for all sequences of strategies $\{\tau^n\}_{n\in \mathbbm N}$ of Bob and for all $n\in \mathbbm N$ we have
\begin{equation}\label{eq:convergence1}
\lambda(\sigma^n,\tau^n)\geq \mathcal J^{(A)}_{\text{cav}} (H(X|Y)) - \mu\left(\frac{1}{n}+\frac{1}{f_n}+\frac{f_n}{n}+g_n+2^{-\frac{1}{2}\left(\frac{n}{f_n}-1\right)h_n\left(\beta g_n-\gamma h_n\right)}\right).
\end{equation}
\end{theorem}
We give an intuitive description of the terms in Equation~\eqref{eq:convergence1} in Discussion \ref{discu1} below.
The formal proof of Theorem~\ref{theorem:finite} is presented in Section~\ref{sec:finite_proof}.

\begin{corollary}\label{cor7}
In the repeated game with leaked randomness source, for each $n\in \mathbbm N$, let $v_n$ denote the max-min value of the $n$-stage game. $v_n$ converges to $\mathcal J^{(A)}_{\text{cav}} (H(X|Y))$ with a rate of at least ${\sqrt{\log n}}/{\sqrt[4]{n}}$. To see this,
let $r,\beta,\gamma,\mu$ be the values in the statement of Theorem~\ref{theorem:finite}, and let $k$ be an arbitrary positive number such that $\beta>k\gamma$ and $kr\leq 1$. Define $f_n=\lceil kr^2(\beta-k\gamma)\sqrt{n}\rceil$, $g_n=r\sqrt{\log n}/\sqrt[4]{n}$, and $h_n=kg_n$. Then, Theorem~\ref{theorem:finite} implies that there exists a sequence of strategies $\{\sigma^n\}_{n\in \mathbbm N}$ such that for all sequences of strategies $\{\tau^n\}_{n\in \mathbbm N}$ of Bob, and for all $n\in \mathbbm N$, we have
$$\lambda(\sigma^n,\tau^n)\geq  \mathcal J^{(A)}_{\text{cav}} (H(X|Y)) - \mathcal{O}\left(g_n\right)=\mathcal J^{(A)}_{\text{cav}} (H(X|Y)) - \mathcal{O}\left(\frac{\sqrt{\log n}}{\sqrt[4]{n}}\right).$$
To see this, observe that $\frac{1}{n}+\frac{1}{f_n}+\frac{f_n}{n}$ is decaying faster than $g_n$. And
$$2^{-\frac{1}{2}(\frac{n}{f_n}-1)h_n(\beta g_n-\gamma h_n)}=\mathcal O\left(2^{-\frac{\sqrt{n}g_n^2}{2r^2}}\right)=\mathcal{O}\left(\frac{1}{\sqrt{n}}\right).$$
\end{corollary}

\begin{discussion}\label{discu1}We explain Equation~\eqref{eq:convergence1} at an intuitive level.
To generate the strategies $\{\sigma^n\}_{n\in \mathbbm N}$ of Theorem~\ref{theorem:finite}, we divide the total $n$ stages almost uniformly into $f_n$ blocks such that the actions of each block (besides the first block) is generated as a function of the randomness source observed during the previous block, and in all stages of the first block, an arbitrary action $a\in \mathcal A$ is played. Therefore, some payoff is lost during the first block; the term $1/f_n$ in Equation~\eqref{eq:convergence1} corresponds with this loss. On the other hand, by dividing the total stages into $f_n$ blocks we get blocks of length at least $n/f_n-1$. This affects the precision of the simulation of the intended distribution of actions from the randomness source observed in previous block, which is reflected in the term 
\begin{align}2^{-\frac{1}{2}(\frac{n}{f_n}-1)h_n(\beta g_n-\gamma h_n)}.\label{eqn:nfm}\end{align}
This equation should be compared with \eqref{eqnNew22}, where the exponent of the simulation error is expressed as the product of three terms: the block length, a term $1-{1}/{\alpha}$, and the entropy difference $H_{\alpha}(X|Y)-H_{\frac{1}{\alpha}}(A)$. 
The term $n/f_n-1$ appears in Equation \eqref{eqn:nfm} as the block length (the lengths of each of the $f_n$ blocks is at least $n/f_n-1$). The sequence $h_n=\alpha-1$ is a proxy for the term 
$1-{1}/{\alpha}$. Finally, considering the last term $H_{\alpha}(X|Y)-H_{\frac{1}{\alpha}}(A)$, we see that larger entropy difference yields better simulation performance. On the other hand, requirement of a large entropy difference restricts the set of action distributions $A$ and results in a payoff loss. The sequence $g_n$ is responsible for this trade-off. Larger $g_n$ results in more loss in payoff (the term $g_n$ in Equation~\ref{eq:convergence1}) but a more accurate simulation (the term $g_n$ in the exponent of the exponential term in Equation~\ref{eq:convergence1}).
\end{discussion}

Next, consider the repeated game with non-causal leaked randomness source (see Remark \ref{remark2}), where the players observe the whole sequence of their corresponding randomness sources before the game starts. We claim the following result:
\begin{theorem}\label{theorem:finite2}
In the repeated game with non-causal leaked randomness source (as described in Remark \ref{remark2}), there exist real numbers $r>0$, $\beta>0$, $\gamma\geq0$ and $\mu\geq 0$ with the following property: for arbitrary sequences of positive numbers $\{g_n\}_{n\in \mathbbm N}$ and $\{h_n\}_{n\in \mathbbm N}$ satisfying $g_n\leq r$ and $h_n\leq 1$, there exists a sequence of strategies $\{\sigma^n\}_{n\in \mathbbm N}$ such that for all sequences of strategies $\{\tau^n\}_{n\in \mathbbm N}$ of Bob and for all $n\in \mathbbm N$ we have
\begin{equation}\label{eq:convergence2}
\lambda(\sigma^n,\tau^n)\geq \mathcal J^{(A)}_{\text{cav}} (H(X|Y)) - \mu\left(\frac{1}{n}+g_n+2^{-\frac{1}{2}nh_n\left(\beta g_n-\gamma h_n\right)}\right).
\end{equation}
\end{theorem}
Proof of Theorem~\ref{theorem:finite2} is given in Section \ref{sec:proof_finite2a}.

\begin{corollary}
In the repeated game with non-causal leaked randomness source, for each $n\in \mathbbm N$, let $v'_n$ denote the max-min value of the $n$-stage game. $v'_n$ converges to $\mathcal J^{(A)}_{\text{cav}} (H(X|Y))$ with a rate of at least ${\sqrt{\log n}}/{\sqrt{n}}$. To see this, let $r,\beta,\gamma,\mu$ be the values in the statement of Theorem~\ref{theorem:finite2}, and let $k$ be an arbitrary positive number such that $\beta>k\gamma$ and $r k\leq 1$. Define $g_n=\min\{r,(\frac{\log n}{k(\beta-k\gamma)n})^{\frac 12}\}$, and $h_n=kg_n$. Then, using similar calculations as in Corollary \ref{cor7}, Theorem~\ref{theorem:finite2} implies that there exists a sequence of strategies $\{\sigma^n\}_{n\in \mathbbm N}$ such that for all sequences of strategies $\{\tau^n\}_{n\in \mathbbm N}$ of Bob, and for all $n\in \mathbbm N$, we have
$$\lambda(\sigma^n,\tau^n)\geq \mathcal J^{(A)}_{\text{cav}} (H(X|Y)) - \mathcal{O}\left(\frac{\sqrt{\log n}}{\sqrt{n}}\right).$$

\end{corollary}

Theorem \ref{theorem:finite} and Theorem \ref{theorem:finite2} provide a convergence rate for general games. However, in some special cases we can derive faster convergence rates for the max-min value of the game. The following theorem provides a special case in which an exponential convergence is obtained.
\begin{theorem}\label{theorem:finite3}
Let $q_A\in \Delta(\mathcal A)$ be an equilibrium strategy for Alice in the one stage game, \emph{i.e.,}
$$q_A\in \underset{p_A\in \Delta(\mathcal A)}{\arg \max}\textrm{ }\min_{b\in \mathcal B}\sum_{a\in \mathcal A}p_A(a)u_{ab}.$$
If $H(X|Y)>H(q_A)$, then, in the repeated game with non-causal leaked randomness source, there exist real numbers $\beta,\gamma>0$, and a sequence of strategies $\{\sigma^n\}_{n\in \mathbbm N}$ such that for all sequences of strategies $\{\tau^n\}_{n\in \mathbbm N}$ of Bob and for all $n\in \mathbbm N$, we have
\begin{equation}\label{eq:convergence3}
\lambda(\sigma^n,\tau^n)\geq \mathcal J^{(A)}_{\text{cav}} (H(X|Y)) - \gamma 2^{-\beta n}.
\end{equation}
\end{theorem}
The proof of Theorem~\ref{theorem:finite3} is provided in Section~\ref{sec:proof_finite3}.

\subsection{A technical tool: simulation of a source from another source}\label{sec:simulation}

To prove the results of Section~\ref{sec:bounded_entropy}, we need a technical tool provided in this section. Here, we study the simulation of a desired single letter source $A\in \mathcal A$ from a given single letter source $X\in \mathcal X$. We assume that $X$ is correlated with a side information $Y\in \mathcal Y$, and we would like the generated source to be almost independent of the side information $Y$. More precisely, we have the following definition:

\begin{definition}
Let $(X,Y)\in \mathcal X\times \mathcal Y$ be distributed according to $p_{XY}$, and $A\in \mathcal A$ be distributed according to $p_A$. We say that the deterministic mapping $f:\mathcal X\to \mathcal A$ simulates $A$ from $X$ with precision $\epsilon$ if we have
$$
\| p_{f(X)Y}-p_Ap_Y\|_{TV} \leq \epsilon,
$$
where $p_{f(X)Y}$ is the joint distribution of $f(X)$ and $Y$.
\end{definition}
According to the above definition, we are interested in a deterministic mapping that simulates $A$ from $X$. However, we utilize the \emph{probabilistic method} and random mappings, as a tool to ultimately prove existence of a suitable deterministic mapping. Therefore, we now define a random mapping and proceed by proving some properties for it. These properties will then lead to the construction of the desired deterministic mapping.

To specify a deterministic mapping $f:\mathcal X\to \mathcal A$, we need to specify the value of $f(x)$ for all $x\in \mathcal X$. To specify a random mapping $F:\mathcal X\to \mathcal A$, we need to specify the joint distribution of the random variables $F(x)$ for $x\in \mathcal X$.
\begin{definition}\label{def:random_mapping}
$F:\mathcal X\to \mathcal A$ is a random mapping constructed as follows: assume that $F(x)$ for different values of $x$ are \emph{i.i.d.\@} according to $p_A(a)$. In other words, given string of symbols $a_x\in\mathcal{A}$ for all $x\in\mathcal{X}$,
$$
\Pr[F(x)=a_x, \forall x\in \mathcal X] =
\prod_{x\in \mathcal X} \Pr[F(x)=a_x]
=\prod_{x\in \mathcal X} p_A(a_x),
$$
The above construction of the random mapping $F$ defines a probability measure $p_F$ on the set of all mappings $f:\mathcal X\to \mathcal A$ denoted by $\mathcal F$.
\end{definition}

\begin{lemma} \label{lemma:simulation}
Let $(X,Y)\in \mathcal X\times \mathcal Y$ be distributed according to $p_{XY}$ and $A\in\mathcal A$ according to $p_A$. Furthermore, let $F$ be the random mapping defined in Definition~\ref{def:random_mapping}. Then, %for all $\alpha\in [1,2]$, we have
\begin{equation} \label{eq:simulation}
\sum_{f\in \mathcal F} p_F(f) \|p_{f(X)Y}-p_{A}p_Y\|_{TV} \leq \min_{\alpha\in [1,2]}\left(2^{-(1-\frac{1}{\alpha})\left (H_{\alpha}(X|Y)-H_{\frac{1}{\alpha}}(A)+2\right)}\right),
\end{equation}
where $p_{f(X)Y}$ is the joint distribution of $f(X)$ and $Y$. Consequently, there exists a deterministic mapping $f:\mathcal X\to \mathcal A$ such that for all $\alpha\in [1,2]$, we have
\begin{equation}\label{eq:simulation2}
\|p_{f(X)Y}-p_{A}p_Y\|_{TV} \leq 2^{-(1-\frac{1}{\alpha})\left (H_{\alpha}(X|Y)-H_{\frac{1}{\alpha}}(A)+2\right)}.
\end{equation}
\end{lemma}
Proof of Lemma~\ref{lemma:simulation} is provided in \ref{sec:proof_simulation}.

While the above inequality ensures the existence of a deterministic mapping $f:\mathcal X\to \mathcal A$ where \eqref{eq:simulation2} holds, it does not provide an explicit mapping $f$. An explicit construction is desirable from an algorithmic perspective. In the following, we address this issue by showing that any randomly chosen mapping $f:\mathcal X\to \mathcal A$ would almost satisfy \eqref{eq:simulation2} with very high probability.

Let $D_{TV}=\|p_{F(X)Y}-p_{A}p_Y\|_{TV}$. The quantity $D_{TV}$ is random because $F$ is random. Thus, random variable $D_{TV}$ is a function of the random variable $F$, \emph{i.e.,} $D_{TV}$ takes value $\|p_{f(X)Y}-p_{A}p_Y\|_{TV}$ with probability $p_F(f)$. Hence, Lemma~\ref{lemma:simulation} implies that for all $\alpha\in [1,2]$,
$$\E[D_{TV}] \leq 2^{-(1-\frac{1}{\alpha})\left (H_{\alpha}(X|Y)-H_{\frac{1}{\alpha}}(A)+2\right)}.$$
We claim the following bound on how $D_{TV}$ concentrates around its expected value.
\begin{proposition}\label{pro:simulation}
For the random variable $D_{TV}$, we have
$$\Pr\Big[\big|D_{TV}-\E[D_{TV}]\big|>t\Big]\leq 2e^{-2t^22^{H_2(X)}}.$$
\end{proposition}
Proof of Proposition~\ref{pro:simulation} is presented in \ref{sec:proof_simulation2}.

One application of Proposition~\ref{pro:simulation} is for simulation of \emph{i.i.d.\@} sequences. Let $(X^n,Y^n)$ be \emph{i.i.d.\@} according to $p_{XY}$, and let $A^n$ be \emph{i.i.d.\@} according to $p_A$. Assume that $H(X|Y)>H(A)$ so that simulation of $A^n$ with arbitrary precision is possible. Let $F:\mathcal X^n\to \mathcal A^n$ be the random mapping of Definition~\ref{def:random_mapping}, where $(X,Y,A)$ is replaced with $(X^n,Y^n,A^n)$. Let us choose $\alpha>1$ such that $H_{\alpha}(X|Y)>H_{\frac{1}{\alpha}}(A)$ (note that such a real number exists since $H(X|Y)>H(A)$, and R\'enyi entropy converges to Shannon entropy as $\alpha$ tends to $1$). Let $\epsilon$ be a positive number such that
$$\epsilon \leq \left(1-\frac{1}{\alpha}\right)\left(H_{\alpha}(X|Y)-H_{\frac{1}{\alpha}}(A)\right),\quad \epsilon< \frac 12 H_2(X).$$
Then, Lemma~\ref{lemma:simulation} implies
\begin{equation}\label{eq:simulation_iid}
\E[D_{TV}]\leq 2^{-\epsilon n},
\end{equation}
where $D_{TV}=\|p_{F(X^n)Y^n}-p_{A^n}p_{Y^n}\|_{TV}$. Furthermore, from Proposition~\ref{pro:simulation}, for $t=2^{-\epsilon n}$, we have
$$\Pr\Big[\big|D_{TV}-\E[D_{TV}]\big|>2^{-\epsilon n}\Big]\leq 2e^{-2^{(H_2(X)-2\epsilon)n}}.$$
The above equation along with Equation~\eqref{eq:simulation_iid} and definition $\delta=H_2(X)-2\epsilon$ implies
$$\Pr[D_{TV}\geq 2\times2^{-\epsilon n}]\leq 2e^{-2^{\delta n}}.$$
In other words, the outcome of the random mapping $F$, with probability at least $1-2e^{-2^{\delta n}}$ (converging double exponentially to $1$) will simulate $A^n$ with precision at most $2\times2^{-\epsilon n}$ (decaying exponentially in $n$).

\iffalse
In the above paragraphs, we showed that for the \emph{i.i.d.\@} case, the outcome of the random mapping $F$ with very high probability is the desired mapping. For the single letter case, we can perform similar analysis. According to Lemma~\ref{lemma:simulation}, there exists a mapping $f^*$ that satisfies Equation~\eqref{eq:simulation2}. The mapping $f^*$ comes out with probability at least $\beta=(p_{min})^{|\mathcal X|}$, where
$$p_{min}=\min_{a\in \mathcal A; p_A(a)>0} p_A(a).$$
Therefore, if we construct independently $m$ random mappings $\{F_i\}_{i=1}^m$ according to Definition~\ref{def:random_mapping}, we will have
$$\Pr[F_i\neq f^*, \forall i=1,\dots,m]\leq (1-\beta)^{m}.$$
In other words, in $m$ trials, a desired mapping will be found with probability at least $1-(1-\beta)^{m}$.
\fi

\subsection{Proof of Theorem~\ref{theorem:finite}}\label{sec:finite_proof}
Let us divide the total stages, $n$, into $f_n$ blocks, where $\{f_n\}_{n\in \mathbbm N}$ is the arbitrary sequence of natural numbers in the statement of the theorem. Let $k_n$ be the remainder of $n$ divided by $f_n$, \emph{i.e.,} $n=\lfloor n/f_n\rfloor f_n+k_n$. Then, the number of stages in each block, $\{N_{n,i}\}_{i=1}^{f_n}$, is computed as follows:
\begin{align} N_{n,i}=\begin{cases}\lfloor n/f_n\rfloor+1&i=1,\dots,k_n\\ \lfloor n/f_n\rfloor&i=k_n+1,\dots,f_n. \end{cases}\label{eqnNnni}\end{align}
In other words, first, all blocks get $\lfloor n/f_n\rfloor$ stages, then, the remaining $k_n$ stages are assigned to the first $k_n$ blocks.

Let $A_i^{N_{n,i}}=(A_{i,1},A_{i,2},\dots,A_{i,N_{n,i}})$ and $B_i^{N_{n,i}}=(B_{i,1},B_{i,2},\dots,B_{i,N_{n,i}})$ denote the sequence of actions played in block $i=1,\dots,f_n$ by Alice and Bob, respectively. Similarly, let $X_i^{N_{n,i}}=(X_{i,1},X_{i,2},\dots,X_{i,N_{n,i}})$ and $Y_i^{N_{n,i}}=(Y_{i,1},Y_{i,2},\dots,Y_{i,N_{n,i}})$ denote the sequence of random sources observed in block $i$ by Alice and Bob, respectively. We generate strategy $\sigma^n$ for Alice as follows: in all stages of the first block, Alice chooses an arbitrary action $a\in \mathcal A$; in each block $i\geq 2$, Alice chooses her action sequence $A_i^{N_{n,i}}$ as a deterministic function of the sequence of random sources observed during the previous block, $X_{i-1}^{N_{n,i-1}}$. Let us denote this deterministic function by $\varphi_i$. Thus, we have
$$A_i^{N_{n,i}}=\varphi_i(X_{i-1}^{N_{n,i-1}}).$$
In order to fulfill the definition of the strategy $\sigma^n$, it suffices to determine the functions $\varphi_i$ for $i=2,\dots,f_n$. We will now determine the functions $\varphi_i$ after presenting some preliminaries.

Considering the definition of the function $\mathcal J^{(A)}_{\text{cav}} (.)$, there exist real number $0\leq r\leq 1$ and pmfs $p_A^{(1)},p_A^{(2)}\in \Delta(\mathcal A)$ such that:
\begin{align}
rU^{(A)}(p_A^{(1)})+(1-r)U^{(A)}(p_A^{(2)})&=\mathcal J^{(A)}_{\text{cav}} (H(X|Y)),\label{eq:pre_1}\\
rH(p_A^{(1)})+(1-r)H(p_A^{(2)}) &\leq H(X|Y).\label{eq:pre_2}
\end{align}
Without loss of generality, we may assume that $H(p_A^{(1)})\geq H(p_A^{(2)})$.
The following lemma claims that we may assume that $r$, $p_A^{(1)}$ and $p_A^{(2)}$ also satisfy the following equations:
\begin{align}
&H(p_A^{(1)})>H(p_A^{(2)}),\label{eq:pre_3}\\
&0<r \leq 1.\label{eq:pre_4}
\end{align}
\begin{lemma} Theorem~\ref{theorem:finite} holds if Equations \eqref{eq:pre_3} and \eqref{eq:pre_4} fail to hold.
\label{lemmaNew1}
\end{lemma}
Proof of the above lemma is given later in Section~\ref{subsectionLemma}.

We identify the value of $r$ in the statement of the theorem as the one given by Equations \eqref{eq:pre_1} and \eqref{eq:pre_2}. The values for $\beta>0$, $\gamma\geq0$ and $\mu\geq 0$ will be identified later. Take an arbitrary sequence $\{g_n\}_{n\in \mathbbm N}$ of positive numbers, as in the statement of the theorem, such that $g_n\leq r$, for all $n\in \mathbbm N$. Let 
$$m_{n,i}=\lfloor N_{n,i}(r-g_n) \rfloor.$$
Moreover, consider an ideal distribution $q_{A_i^{N_{n,i}}}$ defined as follows for $i=2,\dots,f_n$:
\begin{equation}\label{eq:ideal_dist}
q_{A_i^{N_{n,i}}}(a_i^{N_{n,i}})=\prod_{t=1}^{m_{n,i}} p^{(1)}_A(a_{i,t})\prod_{t=m_{n,i}+1}^{N_{n,i}} p^{(2)}_A(a_{i,t}).
\end{equation}
For each $i=2,\dots,f_n$, we choose $\varphi_i$ to be the mapping of Lemma~\ref{lemma:simulation} that simulates $q_{A_i^{N_{n,i}}}$ from $X_{i-1}^{N_{n,i-1}}$; hence, for all $1\leq \alpha\leq 2$, we have
\begin{equation}\label{eq:sim1}
\bigg\| p_{A_i^{N_{n,i}}Y_{i-1}^{N_{n,i-1}}}-q_{A_i^{N_{n,i}}}p_{Y_{i-1}^{N_{n,i-1}}} \bigg\|_{\mathrm{TV}} \leq 2^{-(1-\frac{1}{\alpha})\left (H_{\alpha}\left(X_{i-1}^{N_{n,i-1}}\Big |Y_{i-1}^{N_{n,i-1}}\right)-H_{\frac{1}{\alpha}}\left(q_{A_i^{N_{n,i}}}\right)+2\right)},
\end{equation}
where $p_{A_i^{N_{n,i}}Y_{i-1}^{N_{n,i-1}}}$ is the joint pmf of $A_i^{N_{n,i}}$ and $Y_{i-1}^{N_{n,i-1}}$. Next, note that
\begin{equation}\label{eq:entr1}
H_{1/\alpha}\left(q_{A_i^{N_{n,i}}}\right)=m_{n,i}H_{1/\alpha}\left(p^{(1)}_A\right)+(N_{n,i}-m_{n,i})H_{1/\alpha}\left(p^{(2)}_A\right).
\end{equation}
On the other hand, since $(X_{i-1}^{N_{n,i-1}},Y_{i-1}^{N_{n,i-1}})$ are drawn \emph{i.i.d.\@} from $p_{XY}$, we have
\begin{equation}\label{eq:entr2}
H_{\alpha}\left(X_{i-1}^{N_{n,i-1}}\Big |Y_{i-1}^{N_{n,i-1}}\right)=N_{n,i-1}H_{\alpha}\left(X|Y\right)\geq N_{n,i}H_{\alpha}\left(X|Y\right),
\end{equation}
where we used $N_{n,i-1}\geq N_{n,i}$, which follows from the definition given in Equation \eqref{eqnNnni}.
Moreover, let $r_{n,i}$ be a fractional approximation of $r$ defined as follows
$$r_{n,i}=\frac{m_{n,i}}{N_{n,i}}.$$
Observe that $$r_{n,i}\leq r-g_n.$$
Equations~\eqref{eq:sim1}, \eqref{eq:entr1} and \eqref{eq:entr2} imply
\begin{equation}\label{eq:sim2}
\bigg\| p_{A_i^{N_{n,i}}Y_{i-1}^{N_{n,i-1}}}-q_{A_i^{N_{n,i}}}p_{Y_{i-1}^{N_{n,i-1}}} \bigg\|_{\mathrm{TV}} \leq 2^{-N_{n,i}(1-\frac{1}{\alpha})\left(H_{\alpha}(X|Y)-r_{n,i}H_{1/\alpha}\left(p^{(1)}_A\right)-(1-r_{n,i})H_{1/\alpha}\left(p^{(2)}_A\right)\right)}.
\end{equation}
Using Equations~\eqref{eqnO1}-\eqref{eqnO4}, we bound the exponent of the exponential term in the right-hand side of the above equation as below:
\begin{align}
H_{\alpha}&(X|Y)-r_{n,i}H_{1/\alpha}\left(p^{(1)}_A\right)-(1-r_{n,i})H_{1/\alpha}\left(p^{(2)}_A\right)\notag\\
&\geq H(X|Y)-r_{n,i}H\left(p^{(1)}_A\right)-(1-r_{n,i})H\left(p^{(2)}_A\right)-(\alpha-1)d_1(X|Y)+\notag\\
&\qquad\qquad\left(\frac{1}{\alpha}-1\right)\left(r_{n,i}d_1\left(p^{(1)}_A\right)+(1-r_{n,i})d_1\left(p^{(2)}_A\right)\right)\notag\\
&\qquad\qquad -(\alpha-1)^2d_2(X|Y)-\left(\frac{1}{\alpha}-1\right)^2\left(r_{n,i}d_2\left(p^{(1)}_A\right)+(1-r_{n,i})d_2\left(p^{(2)}_A\right)\right)\nonumber
\\&= H(X|Y)-r_{n,i}H\left(p^{(1)}_A\right)-(1-r_{n,i})H\left(p^{(2)}_A\right)\notag\\
&\qquad\qquad-(\alpha-1)\left(d_1(X|Y)+\frac{1}{\alpha}\left(r_{n,i}d_1\left(p^{(1)}_A\right)+(1-r_{n,i})d_1\left(p^{(2)}_A\right)\right)\right)\notag\\
&\qquad\qquad -(\alpha-1)^2\left(d_2(X|Y)+\frac{1}{\alpha^2}\left(r_{n,i}d_2\left(p^{(1)}_A\right)+(1-r_{n,i})d_2\left(p^{(2)}_A\right)\right)\right)\notag\\
&\geq H(X|Y)-r_{n,i}H\left(p^{(1)}_A\right)-(1-r_{n,i})H\left(p^{(2)}_A\right)\notag\\
&\qquad\qquad-(\alpha-1)\left(d_1(X|Y)+\max \left\{d_1(p_A^{(1)}),d_1(p_A^{(2)})\right\}\right)\notag\\
&\qquad\qquad -(\alpha-1)^2\left(d_2(X|Y)+\max \left\{d_2(p_A^{(1)}),d_2(p_A^{(2)})\right\}\right).
\label{eq:exponent1}
\end{align}
where in \eqref{eq:exponent1} we used the fact that $0\leq r_{n,i}\leq 1$ and $\alpha \geq 1$. 
On the other hand, Equations~\eqref{eq:pre_2} and \eqref{eq:pre_3} along with the fact that $r_{n,i}\leq r-g_n$ imply
\begin{equation}\label{eq:exponent2}
H(X|Y)-r_{n,i}H\left(p^{(1)}_A\right)-(1-r_{n,i})H\left(p^{(2)}_A\right) \geq \beta g_n,
\end{equation}
where $$\beta\triangleq H(p_A^{(1)})-H(p_A^{(2)})>0.$$ Next, let us define
$$\gamma\triangleq 2 \max \left\{d_1(X|Y)+\max \left\{d_1(p_A^{(1)}),d_1(p_A^{(2)})\right\},d_2(X|Y)+\max \left\{d_2(p_A^{(1)}),d_2(p_A^{(2)})\right\} \right\}.$$
Then, Equations \eqref{eq:exponent1} and \eqref{eq:exponent2} imply
\begin{align}
H_{\alpha}(X|Y)-r_{n,i}H_{1/\alpha}\left(p^{(1)}_A\right)-(1-r_{n,i})H_{1/\alpha}\left(p^{(2)}_A\right) &\geq \beta g_n - \frac 12 \gamma (\alpha - 1 +(\alpha-1)^2)\notag\\
&\geq \beta g_n - \gamma (\alpha - 1),\label{eq:exponent3}
\end{align}
where \eqref{eq:exponent3} results from $\alpha \leq 2$. By using \eqref{eq:exponent3} in \eqref{eq:sim2}, and simplifications $1-1/\alpha\geq (\alpha-1)/2$ and $N_{n,i}\geq n/f_n-1$ we obtain
\begin{equation}\label{eq:sim3}
\bigg\| p_{A_i^{N_{n,i}}Y_{i-1}^{N_{n,i-1}}}-q_{A_i^{N_{n,i}}}p_{Y_{i-1}^{N_{n,i-1}}} \bigg\|_{\mathrm{TV}} \leq 2^{-\frac 12 (\frac{n}{f_n}-1)(\alpha-1)\left(\beta g_n-\gamma(\alpha-1)\right)}.
\end{equation}
Next, let $\alpha=1+h_n$, where $\{h_n\}_{n\in \mathbbm N}$ is the arbitrary sequence of positive real numbers in the statement of the theorem. Then, Equation~\eqref{eq:sim3} results in
\begin{equation}\label{eq:sim4}
\bigg\| p_{A_i^{N_{n,i}}Y_{i-1}^{N_{n,i-1}}}-q_{A_i^{N_{n,i}}}p_{Y_{i-1}^{N_{n,i-1}}} \bigg\|_{\mathrm{TV}} \leq 2^{-\frac 12 (\frac{n}{f_n}-1)h_n\left(\beta g_n-\gamma h_n\right)}\triangleq \delta_{n}.
\end{equation}

Now, we need to include the sequence of actions of Bob at the $i$-th block ($B_i^{N_{n,i}}$) into Equation~\eqref{eq:sim4}. To do so, note that $A_i^{N_{n,i}}$ is independent of Alice's actions in all blocks, except for the $i$-th block. This is because the $X$-source is \emph{i.i.d.\@} and $A_i^{N_{n,i}}$ is a function of $X_{i-1}^{N_{n,i-1}}$. Therefore, at $t$-th stage of block number $i$, Bob obtains information about $X_{i-1}^{N_{n,i-1}}$ only through his source $Y_{i-1}^{N_{n,i-1}}$ and prior actions $A_{i-1}^{t-1}$. In other words, $B_{i,t}$ is conditionally independent of $X_{i-1}^{N_{n,i-1}}$ given $Y_{i-1}^{N_{n,i-1}}, A_i^{t-1}, B_i^{t-1}$. Since $A_i^{N_{n,i}}=\varphi_i (X_{i-1}^{N_{n,i-1}})$, $B_{i,t}$ is also conditionally independent of $A_{i,t}, A_{i,t+1}, \cdots, A_{i,N_{n,i}}$ given $Y_{i-1}^{N_{n,i-1}}, A_i^{t-1}, B_i^{t-1}$. Thus,
$$p_{B_i^{N_{n,i}}|A_i^{N_{n,i}}Y_{i-1}^{N_{n,i-1}}}=\prod_{t=1}^{N_{n,i}} p_{B_{i,t}|A_i^{t-1}Y_{i-1}^{N_{n,i-1}}B_i^{t-1}}.$$
Then, utilizing the first property of total variation in Lemma~\ref{lemma:tv_p} for random variables $E=(A_i^{N_{n,i}},Y_{i-1}^{N_{n,i-1}})$ and $F=B_i^{N_{n,i}}$ we conclude from \eqref{eq:sim4} that
\begin{equation*}
\bigg\| p_{A_i^{N_{n,i}}Y_{i-1}^{N_{n,i-1}}B_i^{N_{n,i}}}-q_{A_i^{N_{n,i}}}p_{Y_{i-1}^{N_{n,i-1}}}\prod_{t=1}^{N_{n,i}} p_{B_{i,t}|A_i^{t-1}Y_{i-1}^{N_{n,i-1}}B_i^{t-1}} \bigg\|_{\mathrm{TV}} \leq \delta_n.
\end{equation*}
Next, by utilizing the second property of total variation in Lemma~\ref{lemma:tv_p} for random variables $E=(A_i^{N_{n,i}},B_i^{N_{n,i}})$ and $F=Y_{i-1}^{N_{n,i-1}}$, and replacing $q_{A_i^{N_{n,i}}}$ from Equation~\eqref{eq:ideal_dist} we conclude
\begin{align}
\bigg\| p_{A_i^{N_{n,i}}B_i^{N_{n,i}}}(a^{N_{n,i}},b^{N_{n,i}})&-\prod_{t=1}^{m_{n,i}}p_A^{(1)}(a_t) p_{B_{i,t}|A_i^{t-1}B_i^{t-1}}(b_t|a^{t-1},b^{t-1})\times\notag\\
&\prod_{t=m_{n,i}+1}^{N_{n,i}}p_A^{(2)}(a_t) p_{B_{i,t}|A_i^{t-1}B_i^{t-1}}(b_t|a^{t-1},b^{t-1}) \bigg\|_{\mathrm{TV}} \leq \delta_n\label{eq:sim5}.
\end{align}
In other words, the distribution of the generated actions $A_i^{N_{n,i}}$ is in distance $\delta_n$ from the ideal distribution $q_{A_i^{N_{n,i}}}$. Note that the ideal distribution $q_{A_i^{N_{n,i}}}$ secures payoff $m_{n,i}U^{(A)}(p_A^{(1)})+(N_{n,i}-m_{n,i})U^{(A)}(q_A^{(2)})$ in the $i$-th block. Therefore, in the $i$-th block, the generated strategy $\sigma^n$ secures payoff
$$m_{n,i}U^{(A)}(p_A^{(1)})+(N_{n,i}-m_{n,i})U^{(A)}(p_A^{(2)})-2N_{n,i}\mathsf M\delta_n,$$
where $\mathsf M=\max_{a\in \mathcal A,b\in\mathcal B} |u_{ab}|$. Thus, for arbitrary strategy $\tau^n$ of Bob we have
\begin{align}
\lambda(\sigma^n&,\tau^n)\geq \frac1n\left\{-\mathsf MN_{n,1}+\sum_{i=2}^{f_n} \left(m_{n,i}U^{(A)}(p_A^{(1)})+(N_{n,i}-m_{n,i})U^{(A)}(p_A^{(2)})-2N_{n,i}\mathsf M\delta_n\right)\right\}\notag\\
&\geq \frac1n\left\{-\mathsf MN_{n,1}+\sum_{i=2}^{f_n} N_{n,i}\left(\mathcal J^{(A)}_{\text{cav}} (H(X|Y))-\left(g_n+\frac{1}{N_{n,i}}\right)\Delta U-2\mathsf M\delta_n\right)\right\}\label{eq:utility1}\\
&\geq \mathcal J^{(A)}_{\text{cav}} (H(X|Y))-\Delta U\left(\frac{f_n}{n}+g_n\right)-2\mathsf M\delta_n-\frac{N_{n,1}} {n}\left(\mathsf M+\mathcal J^{(A)}_{\text{cav}} (H(X|Y))\right)\label{eq:utility2}\\
&\geq \mathcal J^{(A)}_{\text{cav}} (H(X|Y))-\Delta U\left(\frac{f_n}{n}+g_n\right)-2\mathsf M\delta_n-2\mathsf M\left(\frac{1}{f_n}+\frac{1}{n}\right),\label{eq:utility3}
\end{align}
where $\Delta U=|U^{(A)}(p_A^{(1)})-U^{(A)}(p_A^{(2)})|$, and inequality~\eqref{eq:utility1} follows from Equation~\eqref{eq:pre_1} and the fact that $|m_{n,i}-rN_{n,i}|\leq g_nN_{n,i}+1$; Equation~\eqref{eq:utility2} is implied by $\sum_{i=1}^{f_n} N_{n,i}=n$, and \eqref{eq:utility3} results from $\mathcal J^{(A)}_{\text{cav}} (H(X|Y))\leq \mathsf M$ and $N_{n,1}\leq n/f_n+1$.

Note that Equation~\eqref{eq:pre_3} implies that $\beta>0$; therefore, by replacing the value of $\delta_n$, and defining $\mu=\max\{2\mathsf M,\Delta U\}$, \eqref{eq:utility3} implies the claim of the theorem.
$~~~~\qquad\qquad\qed$

\subsubsection{Proof of Lemma \ref{lemmaNew1}}\label{subsectionLemma}
We need to consider the case of $r=0$ or $H(p_A^{(1)})=H(p_A^{(2)})$.
\begin{itemize}
\item The case of $r=0$ and $H(p_A^{(2)})=0$: here, $p_A^{(2)}$ is deterministic (it outputs an action $a\in \mathcal A$ with probability $1$), and hence, the trivial strategy of playing $a$ in all stages secures payoff $\mathcal J^{(A)}_{\text{cav}} (H(X|Y))$ for Alice; therefore, in this case, the claim of the theorem holds with $\mu=0$.
\item The case of $r=0$ and $H(p_A^{(2)})>0$: in this case, let $r'=1$, $q_A^{(1)}=p_A^{(2)}$, and let $q_A^{(2)}$ be an arbitrary deterministic pmf. Then, $r'$, $q_A^{(1)}$ and $q_A^{(2)}$ satisfy Equations~\eqref{eq:pre_1}-\eqref{eq:pre_4}. Therefore, we can proceed with the proof of Theorem~\ref{theorem:finite} with these assumptions.
\item If $H(p_A^{(1)})=H(p_A^{(2)})=0$, then Alice can achieve $\mathcal J^{(A)}_{\text{cav}} (H(X|Y))$ by playing a pure action, and the claim of the theorem holds with $\mu=0$. 
\item If $r=1$ and $H(p_A^{(1)})=H(p_A^{(2)})>0$, then, we can change $p_A^{(2)}$ to an arbitrary deterministic pmf so that Equations~\eqref{eq:pre_1}-\eqref{eq:pre_4} hold. Therefore, we can proceed with the proof of Theorem~\ref{theorem:finite} with these assumptions.
\item If $0<r<1$ and $H(p_A^{(1)})=H(p_A^{(2)})>0$, then, we can change $r$ to $r=1$, and $p_A^{(2)}$ to a deterministic pmf such that Equations~\eqref{eq:pre_1}-\eqref{eq:pre_4} hold. This is because $0<r<1$ and $H(p_A^{(1)})=H(p_A^{(2)})>0$ imply that $U^{(A)}(p_A^{(1)})=U^{(A)}(p_A^{(2)})$, since otherwise, by changing $r$ we would get greater value for $\mathcal J^{(A)}_{\text{cav}} (H(X|Y))$, which contradicts the definition of the upper concave envelope. 
\end{itemize}

\subsection{Proof of Theorem~\ref{theorem:finite2}}\label{sec:proof_finite2a}
The proof is similar to the proof of Theorem~\ref{theorem:finite} with few modifications. More specifically, in a repeated game with non-causal leaked randomness source we do not need to divide the total $n$ stages into blocks; instead, we can generate all actions of Alice as a function of the whole randomness source. Let $A^n=(A_{1},A_{2},\dots,A_{n})$ and $B^{n}=(B_{1},B_{2},\dots,B_{n})$ denote the sequences of actions of Alice and Bob, and let $X^{n}=(X_{1},X_{2},\dots,X_{n})$ and $Y^{n}=(Y_{1},Y_{2},\dots,Y_{n})$ denote the sequences of random sources of Alice and Bob, respectively. We generate strategy $\sigma^n$ for Alice such that Alice chooses her action sequence $A^{n}$ as a deterministic function of $X^{n}$, \emph{i.e.,}
$$A^{n}=\varphi_n(X^n).$$
We will now determine the function $\varphi_n$ after presenting some preliminaries.

As stated in the proof of Theorem~\ref{theorem:finite} in Section~\ref{sec:finite_proof}, we assume that there exist real number $r$ and pmfs $p_A^{(1)},p_A^{(2)}\in \Delta(\mathcal A)$ satisfying \eqref{eq:pre_1}-\eqref{eq:pre_4}. Moreover, let
$$m_{n}=\lfloor n(r-g_n) \rfloor,$$
and let $q_{A^{n}}$ be an ideal distribution of actions defined as follows:
\begin{equation}\label{eq:nc_ideal_dist}
q_{A^{n}}(a^n)=\prod_{t=1}^{m_{n}} p^{(1)}_A(a_{t})\prod_{t=m_{n}+1}^{n} p^{(2)}_A(a_{t}).
\end{equation}
We choose $\varphi_n$ to be the mapping of Lemma~\ref{lemma:simulation} that simulates $q_{A^{n}}$ from $X^{n}$; hence, for all $1\leq \alpha\leq 2$ we have
\begin{equation}\label{eq:nc_sim1}
\| p_{A^{n}Y^{n}}-q_{A^{n}}p_{Y^{n}} \|_{\mathrm{TV}} \leq 2^{-(1-\frac{1}{\alpha})\left (H_{\alpha}\left(X^{n} |Y^{n}\right)-H_{\frac{1}{\alpha}}\left(q_{A^{n}}\right)+2\right)}.
\end{equation}
Next, note that $H_{1/\alpha}\left(q_{A^{n}}\right)=m_{n}H_{1/\alpha}\left(p^{(1)}_A\right)+(n-m_{n})H_{1/\alpha}\left(p^{(2)}_A\right)$, and $H_{\alpha}\left(X^n|Y^{n}\right)=nH_{\alpha}\left(X|Y\right)$. Thus, defining $r_{n}=m_{n}/n$, Equation~\ref{eq:nc_sim1} implies
\begin{equation}\label{eq:nc_sim2}
\| p_{A^{n}Y^{n}}-q_{A^{n}}p_{Y^{n}} \|_{\mathrm{TV}} \leq 2^{-n(1-\frac{1}{\alpha})\left(H_{\alpha}(X|Y)-r_{n}H_{1/\alpha}\left(p^{(1)}_A\right)-(1-r_{n})H_{1/\alpha}\left(p^{(2)}_A\right)\right)}.
\end{equation}
A similar argument as the one used to prove Equation~\eqref{eq:exponent3} in Section~\ref{sec:finite_proof} implies
\begin{align}
H_{\alpha}(X|Y)-r_{n}H_{1/\alpha}\left(p^{(1)}_A\right)-(1-r_{n})H_{1/\alpha}\left(p^{(2)}_A\right) \geq \beta g_n-\gamma(\alpha-1),\label{eq:nc_exponent1}
\end{align}
where $\beta=H(p_A^{(1)})-H(p_A^{(2)})$, and 
$$\gamma = 2\max \left\{d_1(X|Y)+\max \left\{d_1(p_A^{(1)}),d_1(p_A^{(2)})\right\},d_2(X|Y)+\max \left\{d_2(p_A^{(1)}),d_2(p_A^{(2)})\right\} \right\}.$$ 
By using \eqref{eq:nc_exponent1} in \eqref{eq:nc_sim2}, and simplification $1-1/\alpha\geq (\alpha-1)/2$, we obtain
\begin{equation}\label{eq:nc_sim3}
\| p_{A^{n}Y^{n}}-q_{A^{n}}p_{Y^{n}} \|_{\mathrm{TV}} \leq 2^{-\frac 12 n(\alpha-1)\left(\beta g_n-\gamma(\alpha-1)\right)}.
\end{equation}
Next, let $\alpha=1+h_n$, where $\{h_n\}_{n\in \mathbbm N}$ is the arbitrary sequence of positive real numbers in the statement of the theorem; hence, Equation~\eqref{eq:nc_sim3} results in
\begin{equation}\label{eq:nc_sim4}
\| p_{A^{n}Y^{n}}-q_{A^{n}}p_{Y^{n}} \|_{\mathrm{TV}} \leq 2^{-\frac 12 nh_n\left(\beta g_n-\gamma h_n\right)}\triangleq \delta'_{n}.
\end{equation}

Now, we need to include the sequence of actions of Bob ($B^n$) into Equation~\eqref{eq:nc_sim4}. Note that at each stage $t$, Bob has access to information $(Y^{t-1},A^{t-1},B^{t-1})$; thus, given an arbitrary strategy $\tau^n$ for Bob, we have
$$p_{B^{n}|A^{n}Y^{n}}=\prod_{t=1}^{n} p_{B_{t}|Y^{n}A^{t-1}B^{t-1}}.$$
Then, using a similar argument as we used in Section~\ref{sec:finite_proof} to prove Equation~\ref{eq:sim5}, the above equation along with Equation~\eqref{eq:nc_sim4} implies
\begin{align}
\bigg \| p_{A^{n}B^{n}}(a^{n},b^{n})-\prod_{t=1}^{m_{n}}&p_A^{(1)}(a_t) p_{B_{t}|A^{t-1}B^{t-1}}(b_t|a^{t-1},b^{t-1})\times\notag\\
&\prod_{t=m_{n}+1}^{n}p_A^{(2)}(a_t) p_{B_{t}|A^{t-1}B^{t-1}}(b_t|a^{t-1},b^{t-1}) \bigg\|_{\mathrm{TV}} \leq \delta'_n\label{eq:nc_sim5}.
\end{align}
In other words, the distribution of the generated actions is in distance $\delta'_n$ from the ideal distribution. Note that the ideal distribution $q_{A^{n}}$ secures payoff $m_{n}U^{(A)}(p_A^{(1)})+(n-m_{n})U^{(A)}(q_A^{(2)})$. Therefore, we have
\begin{align}
\lambda(\sigma^n,\tau^n) &\geq \frac 1n\left\{m_{n}U^{(A)}(p_A^{(1)})+(n-m_{n})U^{(A)}(p_A^{(2)})-2n\mathsf M\delta'_n\right\}\notag\\
&\geq \mathcal J_{cav}(H(X|Y))-\Delta U(g_n+\frac 1n)-2\mathsf M\delta'_n,\notag
\end{align}
where $\mathsf M=\max_{a\in \mathcal A,b\in\mathcal B} |u_{ab}|$, $\Delta U=|U^{(A)}(p_A^{(1)})-U^{(A)}(p_A^{(2)})|$, and the second inequality follows from Equation~\eqref{eq:pre_1} along with the fact that $|m_{n}-rn|\leq g_nn+1$. By replacing $\delta'_n$ and defining $\mu=\max\{2\mathsf M,\Delta U\}$, we obtain the claim of the theorem.

\subsection{Proof of Theorem~\ref{theorem:finite3}}\label{sec:proof_finite3}
The inequality $H(X|Y)> H(q_A)$ along with the fact that $q_A$ is an equilibrium strategy for Alice in the stage game implies that
\begin{equation}\label{eq:spec_starter}
\mathcal J^{(A)}_{\text{cav}} (H(X|Y))=U^{(A)}(q_A).
\end{equation}
If Alice could play \emph{i.i.d.\@} according to $q_A$, she would have secured payoff $U^{(A)}(q_A)$. Our goal is to generate the actions of Alice, $A^n$, as a deterministic function of the randomness source $X^n$ in such a way that at every stage $t$, the distribution of the action $A_t$ is almost $q_A$ and is almost independent of the past observations of Bob.

The strategy $\sigma^n$ is defined as follows: the actions $A^n$ are chosen as a deterministic function of $X^n$, \emph{i.e.,} $A^n=\varphi_n(X^n)$. We will now define the mapping $\varphi_n$. Consider an ideal distribution $q_{A^n}$ defined as below:
\begin{equation}\label{eq:spec_iid_dist}
q_{A^n}(a^n)=\prod_{t=1}^{n}q_A(a_t).
\end{equation}
Let $\varphi_n$ be the mapping of Lemma~\ref{lemma:simulation} that simulates $q_{A^n}$ from $X^n$; hence, for all $1\leq \alpha\leq 2$, we have
\begin{equation}\label{eq:spec_sim1}
\| p_{A^nY^n}-q_{A^n}p_{Y^n} \|_{\mathrm{TV}} \leq 2^{-(1-\frac{1}{\alpha})\left (H_{\alpha}\left(X^n|Y^n\right)-H_{\frac{1}{\alpha}}\left(q_{A^n}\right)+2\right)},
\end{equation}
where $p_{A^nY^n}$ is the joint pmf of $A^n$ and $Y^n$. Note that $(X^n,Y^n)$ are drawn \emph{i.i.d.\@} from $p_{XY}$, and $q_{A^n}$ is \emph{i.i.d.\@} as well, thus, we have
\begin{equation}\label{eq:spec_entr1}
H_{\alpha}(X^n|Y^n)- H_{1/\alpha}(q_{A^n})=n(H_{\alpha}(X|Y)- H_{1/\alpha}(q_{A})).
\end{equation}
Furthermore, let $\beta$ be defined as follows
$$\beta=\sup_{1<\alpha\leq 2}(1-\frac{1}{\alpha})\left(H_{\alpha}(X|Y)- H_{\frac{1}{\alpha}}(q_{A})\right).$$
Note that $\lim_{\alpha\to 1}\left(H_{\alpha}(X|Y)- H_{1/\alpha}(q_{A})\right)=H(X|Y)-H(q_A)>0$; hence, $\beta >0$. Equations~\eqref{eq:spec_sim1} and \eqref{eq:spec_entr1} along with the above definition of $\beta$ imply
\begin{equation}\label{eq:spec_sim2}
\| p_{A^nY^n}-q_{A^n}p_{Y^n} \|_{\mathrm{TV}} \leq 2^{-\beta n}.
\end{equation}
Next, let Bob play an arbitrary strategy $\tau^n$ and let $B^n$ denote the sequence of actions of Bob. At stage $t$, Bob generates $B_t$ as a function of $Y^n$ and his previous observations $A^{t-1}$ and $B^{t-1}$. Hence, we have
$$p_{B^n|A^nY^n}=\prod_{t=1}^n p_{B_t|Y^nA^{t-1}B^{t-1}}.$$
Then, utilizing the first property of total variation in Lemma \ref{lemma:tv_p} for random variables $E=(A^n,Y^n)$ and $F=B^n$, we conclude from \eqref{eq:spec_sim2} that
\begin{equation*}
\left\| p_{A^nY^nB^n}-q_{A^n}p_{Y^n}\prod_{t=1}^n p_{B_{t}|A^{t-1}Y^{n}B^{t-1}} \right\|_{\mathrm{TV}} \leq 2^{-\beta n}.
\end{equation*}
Next, by utilizing the second property of total variation in Lemma~\ref{lemma:tv_p} for random variables $E=(A^n,B^n)$ and $F=Y^n$, and replacing $q_{A^n}$ from Equation~\eqref{eq:spec_iid_dist} we conclude
\begin{align}\label{eq:spec_sim3}
\bigg\| p_{A^nB^n}(a^n,b^n)&-\prod_{t=1}^{n}q_A(a_t) p_{B_{t}|A^{t-1}B^{t-1}}(b_t|a^{t-1},b^{t-1})\bigg\|_{\mathrm{TV}} \leq 2^{-\beta n}.
\end{align}
In other words, the distribution of the generated actions $p_{A^nB^n}$ is in distance $2^{-\beta n}$ from the ideal distribution $\prod_{t=1}^{n}q_A(a_t) p_{B_{t}|A^{t-1}B^{t-1}}(b_t|a^{t-1},b^{t-1})$. Note that the ideal distribution secures payoff $U^{(A)}(q_A)$ for Alice. Therefore, Equation~\eqref{eq:spec_sim3} implies that
$$\lambda(\sigma^n,\tau^n)\geq U^{(A)}(q_A)-2\mathsf M2^{-\beta n},$$
where $\mathsf M=\max_{a\in \mathcal A,b\in\mathcal B} |u_{ab}|$. Note that $\tau^n$ is an arbitrary strategy for Bob, therefore, the above inequality along with \eqref{eq:spec_starter} implies the claim of the theorem.

\section{Approximate Nash equilibria of the repeated game with leaked randomness source}\label{sec:leaked2}
In the repeated game with leaked randomness source defined in Section~\ref{sec:bounded_entropy}, we have forced the players to randomize their actions just by conditioning them to the outcomes of the random sources $X^n$ and $Y^n$. In this setting, Nash equilibria do not necessarily exist (See \cite{Hubacek} and \cite{Budinich}). However, approximate Nash equilibria may exist. The goal of this section is to characterize the set of approximate Nash equilibria achievable by the randomness sources $X^n$ and $Y^n$. To proceed, consider the following definitions.

\begin{definition}
In the $n$ stage repeated game, the strategy profile $(\sigma^n,\tau^n)$ is an $(\epsilon_A,\epsilon_B)$-Nash equilibrium if Alice (resp.\ Bob) can not increase (resp.\ decrease) the expected average payoff (defined in Equation~\eqref{eq:exp_avg}) more than $\epsilon_A$ (resp.\ $\epsilon_B$) by changing her (resp.\ his) strategy unilaterally.
\end{definition}

\begin{definition}
We say $v$ is a $(\epsilon_A,\epsilon_B)$-Nash equilibrium payoff if for arbitrary $\delta>0$ there exists a natural number $n_0$ and a sequence of strategy profiles $\{(\sigma^n,\tau^n)\}_{n\in\mathbbm N}$ such that for all $n\geq n_0$, $(\sigma^n,\tau^n)$ forms a $(\epsilon_A+\delta,\epsilon_B+\delta)$-Nash equilibrium, and $|\lambda(\sigma^n,\tau^n)-v|\leq \delta$.
\end{definition}

\begin{definition}
$(\epsilon_A,\epsilon_B)$-Nash equilibrium is achievable in long run if for all $\delta>0$, there exists a natural number $n_0$ such that for all $n\geq n_0$, in the $n$-stage repeated game, there exists a $(\epsilon_A+\delta,\epsilon_B+\delta)$-Nash equilibrium.
\end{definition}

We will now characterize the set of all approximate Nash equilibria of the repeated game with leaked randomness source. To do so, we first need to comment on the long run security level of the players. As stated in Theorem~\ref{theorem:maxmin_val}, Alice can secure arbitrary payoff $v$ in long run if and only if $v\leq \mathcal J_{cav}^{(A)}(H(X|Y))$, where $\mathcal J_{cav}^{(A)}(.)$ is the upper concave envelope of $\mathcal J^{(A)}(.)$ defined in Definition~\ref{def:security_level}. Using Theorem~\ref{theorem:maxmin_val}, we can derive a similar result from Bob's (the minimizer) point of view. Consider the following definitions:

\begin{definition}\label{def:min_max_value}
Let $v$ be an arbitrary real value:
\begin{itemize}
\item Bob can secure $v$ in the $n$ stage repeated game if there exists a strategy $\tau^n$ for Bob such that for all strategy $\sigma^n$ of Alice we have $\lambda(\sigma^n,\tau^n) \leq v$.
\item Bob can secure $v$ in long run if there exists a sequence of strategies $\{\tau^n\}_{n\in \mathbbm N}$ for Bob such that for all sequences of strategies $\{\sigma^n\}_{n\in \mathbbm N}$ of Alice we have $\limsup_{n\to \infty} \lambda(\sigma^n,\tau^n) \leq v$.
\end{itemize}
\end{definition}

\begin{definition}\label{def:security_level_b}
In a stage game, the security level of mixed action $p_B$ for Bob is denoted by $U^{(B)}(p_B)$, and is defined as follows:
\begin{equation}\label{eq:u_b}
U^{(B)}(p_B)=\max_{a\in\mathcal{A}} \sum_{b\in \mathcal B}p_B(b)u_{ab}.
\end{equation}
Furthermore, the minimum cost that Bob can secure in a stage game, by playing mixed actions of entropy at most $\mathsf h$, is denoted by $\mathcal{J}^{(B)}(\mathsf h)$, and is defined as:
\begin{equation}\label{eq:j_b}
\mathcal{J}^{(B)}(\mathsf h) = \min_{p_B\in \Delta(\mathcal{B}), H(p_B)\leq \mathsf{h}} U^{(B)}(p_B).
\end{equation}
\end{definition}

Next, by replacing the stage payoff $u_{ab}$ with $-u_{ab}$, and hence, considering Bob as the maximizer, we can deduce the following corollary of Theorem~\ref{theorem:maxmin_val}.

\begin{corollary}\label{corollary:minmax_val}
Let $\mathcal J^{(B)}_{\text{vex}} (\mathsf h)$ be the lower convex envelope of $\mathcal{J}^{(B)}(\mathsf h)$ defined in Definition~\ref{def:security_level_b}. In the repeated game with leaked randomness source, Bob can secure $v$ in long run if and only if $v\geq \mathcal J^{(B)}_{\text{vex}} (H(Y|X))$. Furthermore, in $n\in \mathbbm{N}$ stage game, Bob can secure $v$ only if $v\geq \mathcal J^{(B)}_{\text{vex}} (H(Y|X))$.
\end{corollary}

Note that the functions $\mathcal J^{(A)}(\mathtt h)$ and $\mathcal J^{(B)}(\mathtt h)$ are respectively increasing and decreasing in $\mathtt h$. On the other hand, the minimax theorem (\cite{neumann_minimax}) implies that $\mathcal J^{(A)}(+\infty)=\mathcal J^{(B)}(+\infty)$; thus, for arbitrary $\mathtt h$ and $\mathtt h'$, we have $\mathcal J^{(B)}(\mathtt h)\geq \mathcal J^{(A)}(\mathtt h')$. Hence, $\mathcal J^{(B)}_{vex}(H(Y|X))\geq \mathcal J^{(A)}_{cav}(H(X|Y))$.

In the following theorem we characterize the set of achievable $(\epsilon_A,\epsilon_B)$-Nash equilibrium payoffs in terms of the individually secured payoffs $\mathcal J^{(A)}_{cav}(H(X|Y))$ and $\mathcal J^{(B)}_{vex}(H(Y|X))$.

\begin{theorem}\label{theorem:folk}
In the repeated game with leaked randomness source defined in Section	~\ref{sec:bounded_entropy}, $v$ is a $(\epsilon_A,\epsilon_B)$-Nash equilibrium payoff if and only if
\begin{equation}\label{eq:main_folk}
\max \{\underline m, \mathcal J^{(B)}_{vex}(H(Y|X))-\epsilon_A\} \leq v \leq \min\{\overline m, \mathcal J^{(A)}_{cav}(H(X|Y))+\epsilon_B\},
\end{equation}
where $\overline m$ and $\underline m$ are the maximum and minimum entries of the payoff table, respectively ($\overline m=\max_{(a,b)\in \mathcal A\times \mathcal B}u_{ab}$, and $\underline m=\min_{(a,b)\in \mathcal A\times \mathcal B}u_{ab}$).
\end{theorem}
The proof of Theorem~\ref{theorem:folk} is provided in Section~\ref{sec:proof_folk}.
\begin{remark}
If $\epsilon_A=\epsilon_B=0$, the set of payoffs satisfying \eqref{eq:main_folk} is empty unless $H(X|Y)$ and $H(Y|X)$ are large enough such that
\begin{equation}\label{eq:enough_entropy}
J^{(A)}_{cav}(H(X|Y))=J^{(B)}_{vex}(H(Y|X))=v^*,
\end{equation}
where $v^*=\max_{p_A\in \Delta(\mathcal A)}\min_{p_B\in \Delta(\mathcal B)} \sum_{a\in \mathcal A, b\in \mathcal B}p_A(a)p_B(b)u_{ab}$. In this case, if \eqref{eq:enough_entropy} holds, the only equilibrium payoff is $v^*$, \emph{i.e.,} the max-min value of the stage game. This particular result coincides with the result of "Folk-Theorem" for two-player zero-sum repeated games in which players can freely randomize their actions.
\end{remark}
We can refine Theorem~\ref{theorem:folk} to characterize the set of $\epsilon_A$ and $\epsilon_B$ for which $(\epsilon_A,\epsilon_B)$-Nash equilibrium is achievable in long run. Let $\epsilon_A$ and $\epsilon_B$ be arbitrary positive numbers. If there exists a $v$ satisfying Equation~\eqref{eq:main_folk}, then, Theorem~\ref{theorem:folk} implies that $(\epsilon_A,\epsilon_B)$-Nash equilibrium is achievable. On the other hand, let $(\sigma^n,\tau^n)$ form an $(\epsilon_A,\epsilon_B)$-Nash equilibrium; then, Theorem~\ref{theorem:folk} implies that $v=\lambda(\sigma^n,\tau^n)$ satisfies \eqref{eq:main_folk}. In other words, $(\epsilon_A,\epsilon_B)$-Nash equilibrium is achievable if and only if there exists a real number $v$ satisfying \eqref{eq:main_folk}. Therefore, by removing $v$ from Equation~\eqref{eq:main_folk}, and rewriting it in terms of $\epsilon_A$ and $\epsilon_B$, we conclude the following corollary of Theorem~\ref{theorem:folk}.

\begin{corollary}\label{corollary:main_theorem}
In the repeated game with leaked randomness source defined in Section	~\ref{sec:bounded_entropy}, $(\epsilon_A,\epsilon_B)$-Nash equilibrium is achievable in long run if and only if
\begin{equation}\label{eq:main_theorem}
\epsilon_A+\epsilon_B \geq \mathcal J^{(B)}_{vex}(H(Y|X)) - \mathcal J^{(A)}_{cav}(H(X|Y)).
\end{equation}
\end{corollary}

\subsection{Proof of Theorem~\ref{theorem:folk}}\label{sec:proof_folk}
We prove that inequality \eqref{eq:main_folk} is both necessary and sufficient for $v$ to be a $(\epsilon_A,\epsilon_B)$-Nash equilibrium payoff.

\textbf{Inequality \eqref{eq:main_folk} is necessary:} In the $n$ stage repeated game, let $\sigma^n$ and $\tau^n$ be arbitrary strategies for Alice and Bob generating an $(\epsilon'_A,\epsilon'_B)$-Nash equilibrium. According to Corollary~\ref{corollary:minmax_val}, given the strategy $\tau^n$ for Bob, there exists a strategy $\sigma^{*n}$ for Alice such that $\lambda(\sigma^{*n},\tau^n)\geq \mathcal J^{(B)}_{vex}(H(Y|X))$. Hence,
$$\lambda(\sigma^{*n},\tau^n)- \lambda(\sigma^n,\tau^n)\geq \mathcal J^{(B)}_{vex}(H(Y|X))-\lambda(\sigma^n,\tau^n). $$
But $(\sigma^n,\tau^n)$ is a $(\epsilon'_A,\epsilon'_B)$-Nash equilibrium, thus, we should have
\begin{equation}\label{eq:proof_1}
\epsilon'_A\geq\mathcal J^{(B)}_{vex}(H(Y|X))-\lambda(\sigma^n,\tau^n).
\end{equation}
Similarly, Theorem~\ref{theorem:maxmin_val} implies that given the strategy $\sigma^n$ for Alice, there exists a strategy $\tau^{*n}$ for Bob such that $\lambda(\sigma^n,\tau^{*n})\leq \mathcal J^{(A)}_{cav}(H(X|Y))$. Hence,
$$\lambda(\sigma^n,\tau^n)- \lambda(\sigma^n,\tau^{*n})\geq \lambda(\sigma^n,\tau^n)-\mathcal J^{(A)}_{cav}(H(X|Y)). $$
But $(\sigma^n,\tau^n)$ is a $(\epsilon'_A,\epsilon'_B)$-Nash equilibrium, thus, we should have
\begin{equation}\label{eq:proof_2}
\epsilon'_B\geq\lambda(\sigma^n,\tau^n)-\mathcal J^{(A)}_{cav}(H(X|Y)).
\end{equation}
On the other hand, since $\lambda(\sigma^n,\tau^n)$ is a convex combination of the entries of the payoff table, we have $\underline m\leq\lambda(\sigma^n,\tau^n)\leq \overline m$; this fact along with Equations~\eqref{eq:proof_1} and \eqref{eq:proof_2} implies that
$$
\max \{\underline m, \mathcal J^{(B)}_{vex}(H(Y|X))-\epsilon'_A\} \leq \lambda(\sigma^n,\tau^n) \leq \min\{\overline m, \mathcal J^{(A)}_{cav}(H(X|Y))+\epsilon'_B\}.
$$
For $v$ to be achievable, the above relation should be satisfied for $\lambda(\sigma^n,\tau^n)$, $\epsilon'_A$ and $\epsilon'_B$ arbitrarily close to $v$, $\epsilon_A$ and $\epsilon_B$, respectively. Thus, Equation~\eqref{eq:main_folk} must hold.

\textbf{Inequality \eqref{eq:main_folk} is sufficient:}
Let $v$, $\epsilon_A\geq 0$ and $\epsilon_B\geq 0$ be real numbers satisfying \eqref{eq:main_folk}. Equation~\eqref{eq:main_folk} implies that $\underline{m}\leq v\leq \overline{m}$; hence, $v$ can be expressed as a convex combination of the entries of the payoff table; \emph{i.e.,} there exist action profiles $(a_1,b_1),(a_2,b_2),\dots,(a_r,b_r)\in \mathcal A\times \mathcal B$, and non-negative numbers $\alpha_1,\alpha_2,\dots,\alpha_r$ summing to one such that
$$v=\sum_{i=1}^r \alpha_i u_{a_ib_i}.$$
Let us approximate each $\alpha_i$ by a rational number $k_i/K$ such that $\sum_{i=1}^r k_i=K$; for arbitrary $\delta>0$, we can choose $K$ large enough such that
\begin{equation} \label{eq:approximate_1}
|v-\hat{v}|\leq \delta,
\end{equation}
where $\hat{v}=\sum_{i=1}^r \frac{k_i}{K} u_{a_ib_i}$. We take $K$ so large that not only inequality \eqref{eq:approximate_1} is satisfied, but also there exist strategies $\sigma^{*K}$ and $\tau^{*K}$ such that in the $K$-stage repeated game, $\sigma^{*K}$ secures expected average payoff of $\mathcal J^{(A)}_{cav}(H(X|Y))-\delta$ for Alice, and $\tau^{*K}$ secures $\mathcal J^{(B)}_{vex}(H(Y|X))+\delta$ for Bob (Such strategies $\sigma^{*K}$ and $\tau^{*K}$ exist since in long run, Alice can secure $\mathcal J^{(A)}_{cav}(H(X|Y))$, and Bob can secure $\mathcal J^{(B)}_{vex}(H(Y|X))$).

Now, we are ready to construct the desired approximate Nash equilibrium $(\sigma^n,\tau^n)$. Let the total stages of the game be of the form $n=NK$, and let us divide the total $n$ stages into $N$ blocks of length $K$. The value of $N$ will be set in the sequel. In each block, Alice and Bob cycle through the action profiles $(a_1,b_1),\dots,(a_r,b_r)$ such that each action profile $(a_i,b_i)$ is repeated in $k_i$ stages. Note that the actions $(a_1,b_1),\dots,(a_r,b_r)$ are deterministic, thus, each player can monitor the actions of the other player to see if he/she is still following the rule or not. If Alice (resp. Bob) deviates the rule, then, in the upcoming blocks, Bob (resp. Alice) plays according to the strategy $\tau^{*K}$ (resp. $\sigma^{*K}$) to secure payoff $\mathcal J^{(B)}_{vex}(H(Y|X))+\delta$ (resp. $\mathcal J^{(A)}_{cav}(H(X|Y))-\delta$).

When Alice and Bob both play according to respective strategies $\sigma^n$ and $\tau^n$, the expected average payoff equals $\hat{v}$, \emph{i.e.,}
\begin{equation} \label{eq:payoff}
\lambda(\sigma^n,\tau^n)=\hat{v}.
\end{equation}

Next, we show that the strategy profile $(\sigma^n,\tau^n)$ forms the desired approximate Nash equilibrium. Let Alice deviate from strategy $\sigma^n$, and play an arbitrary strategy $\sigma'^n$. Furthermore, let the deviation be detected by Bob at block $j\in\{1,2,\dots,N\}$. The expected average payoff will be $\hat{v}$ in the blocks before the $j$-th block, and the payoff of the blocks after the $j$-th block (where Bob plays $\tau^{*K}$) will be at most $\mathcal J^{(B)}_{vex}(H(Y|X))+\delta$. In the $j$-th block, Alice could get at most $\mathsf M=\max_{(a,b)\in \mathcal A\times \mathcal B} |u_{ab}|$, thus,
\begin{equation}\label{eq:main_inequality}
\lambda(\sigma'^n,\tau^n) \leq \frac{j-1}{N} \hat{v}+\frac {\mathsf M}{N} + \frac{N-j}{N} (\mathcal J^{(B)}_{vex}(H(Y|X))+\delta).
\end{equation}
Equation~\eqref{eq:main_inequality} along-with Equation~\eqref{eq:payoff} implies:
\begin{align}
\lambda_n(\sigma'^n,\tau^n)-\lambda(\sigma^n,\tau^n)&\leq \frac {1}{N}(\mathsf M-\hat v) + \frac{N-j}{N} (\mathcal J^{(B)}_{vex}(H(Y|X))-\hat v+\delta)\notag\\
&\leq \frac{2\mathsf M}{N}+\frac{N-j}{N} (\mathcal J^{(B)}_{vex}(H(Y|X))-v+2\delta),\label{eq:improve_0}
\end{align}
where \eqref{eq:improve_0} follows from Equation~\eqref{eq:approximate_1}, and the fact that $\hat v\geq -\mathsf M$. On the other hand, Equation~\ref{eq:main_folk} implies:
\begin{equation}\label{eq:folk_fact}
\mathcal J^{(B)}_{vex}(H(Y|X))-v \leq \epsilon_A.
\end{equation}
Equations~\eqref{eq:improve_0} and \eqref{eq:folk_fact} imply:
\begin{align}
\lambda_n(\sigma'^n,\tau^n)-\lambda(\sigma^n,\tau^n) \leq \epsilon_A + \frac{2\mathsf M}{N} + 2\delta.\label{eq:improve_1}
\end{align}

By a similar argument we can also show that for arbitrary strategy $\tau'^n$ for Bob we have
\begin{equation}\label{eq:improve_2}
\lambda(\sigma^n,\tau^n)-\lambda(\sigma^n,\tau'^n)\leq \epsilon_B + \frac{2\mathsf M}{N} + 2\delta.
\end{equation}
Inequalities \eqref{eq:improve_1} and \eqref{eq:improve_2} imply that the strategy profile $(\sigma^n,\tau^n)$ forms a $(\epsilon_A+2\mathsf M/N+2\delta,\epsilon_B+2\mathsf M/N+2\delta)$-Nash equilibrium with expected average payoff $\hat v$. We can choose $\delta$ small enough and $N$ large enough to make $2\mathsf M/N+2\delta$ as small as desired, and hence, $\hat v$ as close to $v$ as desired (according to \eqref{eq:approximate_1}). Thus, $v$ is an $(\epsilon_A,\epsilon_B)$-Nash equilibrium payoff.

\subsection{Approximate Nash equilibria achieved by autonomous strategies}\label{sec:autonomous}
We call a strategy an autonomous strategy if the action of each stage is indifferent about the actions of the opponent in the previous stages. Formally, in the $n$ stage repeated game with leaked randomness sources defined in Section~\ref{sec:bounded_entropy}, strategies $\sigma^n=(\sigma_1,\dots,\sigma_n)$ and $\tau^n=(\tau_1,\dots,\tau_n)$ are autonomous if for arbitrary $t\in\{1,\dots,n\}$ and arbitrary histories $a^{t-1},\tilde{a}^{t-1}\in\mathcal A^{t-1}$, $b^{t-1},\tilde{b}^{t-1}\in \mathcal B^{t-1}$, $x^t\in\mathcal X^t$ and $y^t\in \mathcal Y^t$ we have
$$
\sigma_t(x^t,a^{t-1},b^{t-1})=\sigma_t(x^t,a^{t-1},\tilde{b}^{t-1}), \quad \tau_t(y^t,a^{t-1},b^{t-1})=\tau_t(y^t,\tilde{a}^{t-1},b^{t-1}).
$$

Autonomous strategies are sufficient for construction of a Nash equilibrium for two-player zero-sum repeated games, where players can freely randomize their actions. Furthermore, in the repeated game with leaked randomness source, the maximum securable payoff of Alice (the max-min payoff) can be secured by an autonomous strategy (see \cite[Section 3.3]{bounded_entropy}). Therefore, we are also interested in the set of approximate Nash equilibria achievable by the class of autonomous strategies.

In this section, we characterize the set of approximate Nash equilibria achievable by autonomous strategies in a simplified version of the repeated game with leaked randomness source. In the simplified version, we assume that the randomness sources $X^n$ and $Y^n$ are independent, \emph{i.e.,} $p_{XY}=p_Xp_Y$, thus, we call it the repeated game with independent randomness sources. It will turn out that the set of approximate Nash equilibria achievable by autonomous strategies is strictly smaller than the set of all achievable approximate Nash equilibria in Corollary~\ref{corollary:main_theorem}. To proceed we need the following definition.

\begin{definition}
$(\epsilon_A,\epsilon_B)$-Nash equilibrium is achievable by autonomous strategies if for all $\delta>0$ there exists a natural number $n_0$ and a sequence of autonomous strategy profiles $\{(\sigma^n,\tau^n)\}_{n\in \mathbbm N}$ such that for all $n\geq n_0$, $(\sigma^n,\tau^n)$ forms a $(\epsilon_A+\delta,\epsilon_B+\delta)$-Nash equilibrium in the $n$ stage repeated game.
\end{definition}

\begin{theorem}\label{theorem:autonomous}
In the repeated game with independent randomness sources, $(\epsilon_A,\epsilon_B)$-Nash equilibrium is achievable by autonomous strategies if and only if there exist random variables $A\in\mathcal{A}$, $B\in \mathcal{B}$ and $Q\in\{0,1,2,3\}$ such that $p_{ABQ}(a,b,q)=p_Q(q)p_{A|Q}(a|q)p_{B|Q}(b|q)$ and
\begin{align*}
&H(A|Q) \leq H(X), \quad H(B|Q)\leq H(Y), \\
&g_A(A,B|Q)\leq\epsilon_A, \quad g_B(A,B|Q) \leq \epsilon_B,
\end{align*}
where $g_A(A,B|Q)$ and $g_B(A,B|Q)$ are defined as follows
\begin{align*}
g_A(A,B|Q) &=\sum_{q=0}^3 p_Q(q)\left[ \max_{a\in \mathcal{A}} \E[u_{aB}|Q=q]- \E[u_{AB}|Q=q]\right], \\
g_B(A,B|Q) &=\sum_{q=0}^3 p_Q(q)\left[ \E[u_{AB}|Q=q] - \min_{b\in \mathcal{B}} \E[u_{Ab}|Q=q]\right].
\end{align*}
\end{theorem}
Proof of Theorem~\ref{theorem:autonomous} is provided in \ref{sec:proof_autonomous}.

\begin{example}\label{example_1}
Consider a repeated game with independent randomness sources $X^n$ and $Y^n$ such that $H(X)=0$, and $H(Y)=1$. The sets of actions of Alice and Bob are $\mathcal A=\mathcal B =\{0,1\}$, and the payoff table is as follows:
$$u_{00}=u_{11}=1,\quad u_{01}=u_{10}=-1.$$
Since $H(X)=0$, Alice must play deterministic actions by which she can secure at most $-1$; hence, $\mathcal J^{(A)}_{cav}(H(X|Y))=-1$. On the other hand, Bob has access to one bit randomness per stage, thus, in each stage, he can play according to the max-min strategy of the one shot game and secure $0$, hence, $\mathcal J^{(B)}_{vex}(H(Y|X))=0$. Consequently, Corollary~\ref{corollary:main_theorem} implies that $(\epsilon_A,\epsilon_B)$-Nash equilibrium is achievable if and only if $\epsilon_A+\epsilon_B\geq 1$. Hence, $(1/2,1/2)$-Nash equilibrium is achievable. It is straightforward to check that $\epsilon_A=1/2$ and $\epsilon_B=1/2$ does not satisfy the conditions of Theorem~\ref{theorem:autonomous}; therefore, in the repeated game of this example, $(1/2,1/2)$-Nash equilibrium is not achievable by autonomous strategies.
\end{example}
\begin{remark}
In the repeated game of Example~\ref{example_1}, the set of approximate Nash equilibria achievable by autonomous strategies is strictly smaller than the set of approximate Nash equilibria achievable by arbitrary strategies. Therefore, for achieving approximate equilibria of the repeated games with leaked randomness source, autonomous strategies are not sufficient.
\end{remark}

\appendix

\setcounter{equation}{0}
\renewcommand{\theequation}{\Alph{section}.\arabic{equation}}
%\numberwithin{equation}{section}
\renewcommand{\thesection}{Appendix \Alph{section}}

\setcounter{equation}{0}
\section{Proof of Lemma~\ref{lemma:simulation}}\label{sec:proof_simulation}
For arbitrary $\alpha\in[1,2]$ we have
\begin{align}
\sum_{f\in \mathcal F} &p_F(f) \|p_{f(X)Y}-p_Ap_Y\|_{TV} \notag \\
&= \frac12\sum_{f\in \mathcal F} \left(p_F(f) \sum_{a\in \mathcal A, y\in \mathcal Y} \bigg|\Pr[f(X)=a,Y=y]-p_A(a)P_Y(y)\bigg|\right) \notag \\
&=\frac12\sum_{y\in \mathcal Y} \left ( p_Y(y) \sum_{f\in \mathcal F} \left (p_F(f) \sum_{a\in \mathcal A}\left |\bigg(\sum_{x\in \mathcal X}p_{X|Y}(x|y) \mathbbm{1}(f(x)=a)\bigg)-p_A(a) \right |\right)\right)\label{eq:sim_proof_1}\\
&=\frac12\sum_{y\in \mathcal Y}\left( p_Y(y) \sum_{a\in \mathcal A}\sum_{f\in \mathcal F}\left( p_F(f)\sqrt[\alpha]{\left |\sum_{x\in \mathcal X}p_{X|Y}(x|y)\bigg( \mathbbm{1}(f(x)=a)-p_A(a)\bigg) \right |^{\alpha}}\right)\right)\label{eq:sim_proof_3}\\
&\leq\frac12\sum_{y\in \mathcal Y} \left(p_Y(y) \sum_{a\in \mathcal A}\sqrt[\alpha]{\sum_{f\in \mathcal F}\left( p_F(f)\left |\sum_{x\in \mathcal X}p_{X|Y}(x|y)\bigg( \mathbbm{1}(f(x)=a)-p_A(a)\bigg) \right |^{\alpha}\right)}\right),\label{eq:sim_proof_2}
\end{align}
where $\mathbbm{1}(.)$ is the indicator function, and Equation \eqref{eq:sim_proof_1} follows from
$$\Pr[f(X)=a,Y=y]=p_Y(y)\Pr[f(X)=a|Y=y]=p_Y(Y)\sum_{x\in \mathcal X}p_{X|Y}(x|y)\mathbbm{1}(f(x)=a),$$
and reordering the summations. Equation~\eqref{eq:sim_proof_3} follows from $\sqrt[1/\alpha]{\beta^{\alpha}}=\beta$, and $\sum_{x\in \mathcal X} p_{X|Y}(x|y)=1$. Inequality~\eqref{eq:sim_proof_2} is implied by utilizing the Jensen's inequality for concave function $\sqrt[\alpha]{\cdot}$.

Next, we claim that for arbitrary $y\in \mathcal Y$ and $a\in \mathcal A$,
\begin{equation}\label{eq:sim_claim}
\sum_{f\in \mathcal F} \left(p_F(f)\left |\sum_{x\in \mathcal X}p_{X|Y}(x|y)\bigg( \mathbbm{1}(f(x)=a)-p_A(a) \bigg)\right |^{\alpha} \right)\leq 2^{2-\alpha}p_A(a)\sum_{x\in\mathcal X} p_{X|Y}(x|y)^{\alpha}.
\end{equation}
Therefore, Equations~\eqref{eq:sim_proof_2} and \eqref{eq:sim_claim} imply
\begin{align}
\sum_{f\in \mathcal F} p_F(f) \|p_{f(X)Y}-p_Ap_Y\|_{TV} &\leq \frac12\sum_{y\in \mathcal Y}\left( p_Y(y) \sum_{a\in \mathcal A} \left(2^{\frac{2}{\alpha}-1}p_A(a)^{\frac{1}{\alpha}}\left(\sum_{x\in\mathcal X} p_{X|Y}(x|y)^{\alpha}\right)^{\frac{1}{\alpha}}\right)\right)\notag\\
& = 2^{2(\frac{1}{\alpha}-1)} \left(\sum_{a\in \mathcal A} p_A(a)^{\frac{1}{\alpha}}\right) \sum_{y\in \mathcal Y}\left(p_Y(y)\left(\sum_{x\in\mathcal X} p_{X|Y}(x|y)^{\alpha}\right)^{\frac{1}{\alpha}}\right) \notag \\
&=2^{2(\frac{1}{\alpha}-1)}2^{\log\left(\sum_{a\in \mathcal A} p_A(a)^{\frac{1}{\alpha}}\right)}2^{\log\left(\sum_{y\in \mathcal Y}\left(p_Y(y)\left(\sum_{x\in\mathcal X} p_{X|Y}(x|y)^{\alpha}\right)^\frac{1}{\alpha}\right)\right)}\notag\\
&=2^{-(1-\frac{1}{\alpha})\left(H_{\alpha}(X|Y)-H_{\frac{1}{\alpha}}(A)+2\right)}. \notag
\end{align}

The above equations fulfills the proof. Thus, we only need to prove the claim of Equation \eqref{eq:sim_claim}. Instead of proving Equation~\eqref{eq:sim_claim}, we prove Equation~\eqref{eq:sim_claim2} which is obtained by replacing $p_{X|Y}(x|y)$ with an arbitrary real function $g:\mathcal X\to \bR$:
\begin{equation}\label{eq:sim_claim2}
\sum_{f\in \mathcal F} \left(p_F(f)\left |\sum_{x\in \mathcal X}g(x)\bigg( \mathbbm{1}(f(x)=a)-p_A(a) \bigg)\right |^{\alpha} \right)\leq 2^{2-\alpha}p_A(a)\sum_{x\in\mathcal X} |g(x)|^{\alpha}.
\end{equation}
In order to interpret the above inequality, let us define $\sigma$-finite measure spaces $(\mathcal X,\Sigma_{\mathcal X},\mu_{\mathcal X})$ and $(\mathcal F,\Sigma_{\mathcal F},\mu_{\mathcal F})$, where for all $x\in \mathcal X$, $\mu_{\mathcal X}(x)=1$, and for all $f\in \mathcal F$, $\mu_{\mathcal F}(f)=p_F(f)$. Furthermore, let $T:\mathcal G_{\mathcal X}\to \mathcal G_{\mathcal F}$ be a linear operator that maps $\mathcal G_{\mathcal X}$ (the set of real valued functions on $\mathcal X$) to $\mathcal G_{\mathcal F}$ (the set of real valued functions on $\mathcal F$) and is defined as below
\begin{equation}\label{eq:operator_T_def}
(Tg)(f)=\sum_{x\in \mathcal X}g(x)\big( \mathbbm{1}(f(x)=a)-p_A(a)\big), \forall g\in \mathcal G_{\mathcal X}\textrm{ and } f\in \mathcal F.
\end{equation}
Moreover, consider the following definition:
\begin{definition}
Let $(\mathcal Y, \Sigma_{\mathcal Y}, \mu_{\mathcal Y})$ and $(\mathcal Z, \Sigma_{\mathcal Z}, \mu_{\mathcal Z})$ be $\sigma$-finite measure spaces, and $h:\mathcal Y\to \mathbbm R$ be a real function on the measure space $(\mathcal Y, \Sigma_{\mathcal Y}, \mu_{\mathcal Y})$. For arbitrary $\beta_1>0$, the $\beta_1$-norm of $h$ is denoted by $\|h\|_{L^{\beta_1}(\mu_{\mathcal Y})}$ and is defined as below:
$$\|h\|_{L^{\beta_1}(\mu_{\mathcal Y})}=\left(\int_{\mathcal Y} |h(y)|^{\beta_1}d\mu_{\mathcal Y}\right)^{\frac{1}{\beta_1}}.$$
$L^{\beta_1}(\mu_{\mathcal Y})$ denotes the set of real functions $h:\mathcal Y\to \mathbbm R$ with bounded $\beta_1$-norm, \emph{i.e.,} $\|h\|_{L^{\beta_1}(\mu_{\mathcal Y})}<\infty$. For arbitrary $\beta_2>0$, let $M:L^{\beta_1}(\mu_{\mathcal Y}) \to L^{\beta_2}(\mu_{\mathcal Z})$ be an operator that maps the real functions on the measure space $(\mathcal Y, \Sigma_{\mathcal Y}, \mu_{\mathcal Y})$ to the real functions on the measure space $(\mathcal Z, \Sigma_{\mathcal Z}, \mu_{\mathcal Z})$. $\|M\|_{L^{\beta_1}(\mu_{\mathcal Y})\to L^{\beta_2}(\mu_{\mathcal Z})}$ denotes the operator norm of $M$ defined as follows:
$$\|M\|_{L^{\beta_1}(\mu_{\mathcal Y})\to L^{\beta_2}(\mu_{\mathcal Z})}=\inf\{c\geq 0:\|Mh\|_{L^{\beta_2}(\mu_{\mathcal Z})}\leq c \|h\|_{L^{\beta_1}(\mu_{\mathcal Y})}, \forall h\in L^{\beta_1}(\mu_{\mathcal Y}) \}.$$
\end{definition}

Using the above definition and Equation~\eqref{eq:operator_T_def}, we can rewrite Equation~\eqref{eq:sim_claim2} as follows
\begin{equation}\label{eq:sim_claim3}
\|T\|_{L^{\alpha}(\mu_{\mathcal X})\to L^{\alpha}(\mu_{\mathcal F})} \leq 2^{\frac{2}{\alpha}-1}p_A(a)^{\frac{1}{\alpha}},
\end{equation}

In order to prove Equation~\eqref{eq:sim_claim3}, it suffices to prove it for the special cases $\alpha=1$ and $\alpha=2$, then, the general form with arbitrary $\alpha \in [1,2]$ will be concluded from the well-known Riesz-Thorin interpolation theorem.
\begin{theorem}[Riesz-Thorin Interpolation Theorem]
Let $(\Omega_1,\Sigma_1,\mu_1)$ and $(\Omega_2,\Sigma_2,\mu_2)$ be arbitrary $\sigma$-finite measure spaces. Suppose $0\leq r_0 \leq r_1 \leq \infty$, $0\leq s_0 \leq s_1 \leq \infty$, and let $T$ be an arbitrary linear operator that maps $L^{r_0}(\mu_1)$ and $L^{r_1}(\mu_1)$ boundedly into $L^{s_0}(\mu_2)$ and $L^{s_1}(\mu_2)$, respectively. For arbitrary $0\leq \theta \leq 1$, let $1/r_{\theta}=(1-\theta)/r_0+\theta/r_1$ and $1/s_{\theta}=(1-\theta)/s_0+\theta/s_1$, then, $T$ maps $L^{r_{\theta}}(\mu_1)$ boundedly into $L^{s_{\theta}}(\mu_2)$ and satisfies the operator norm estimate
$$\|T\|_{L^{r_{\theta}}(\mu_1)\to L^{s_{\theta}}(\mu_2)} \leq \|T\|^{1-\theta}_{L^{r_0}(\mu_1)\to L^{s_0}(\mu_2)}\|T\|^{\theta}_{L^{r_1}(\mu_1)\to L^{s_1}(\mu_2)}.$$
\end{theorem}

We complete the proof by proving Equation~\eqref{eq:sim_claim3}, or equivalently Equation~\eqref{eq:sim_claim2}, for $\alpha=1$ and $\alpha=2$. For $\alpha=1$, we have

\begin{align}
\sum_{f\in \mathcal F} &\left( p_F(f)\left |\sum_{x\in \mathcal X}g(x)\bigg( \mathbbm{1}(f(x)=a)-p_A(a) \bigg)\right|\right) \notag \\
&\leq \sum_{f\in \mathcal F} \left(p_F(f)\sum_{x\in \mathcal X}|g(x)|\big |\mathbbm{1}(f(x)=a)-p_A(a) \big|\right)\notag\\
&= \sum_{x\in \mathcal X}\left(|g(x)|\sum_{f\in \mathcal F}p_F(f)\big |\mathbbm{1}(f(x)=a)-p_A(a) \big|\right)\notag\\
&= \sum_{x\in \mathcal X}\bigg(|g(x)|\bigg(p_A(a)|1-p_A(a)|+(1-p_A(a))|0-p_A(a)|)\bigg)\bigg)\label{eq:sim_proof_prob}\\
&= 2p_A(a)(1-p_A(a))\left(\sum_{x\in \mathcal X}|g(x)|\right)\notag\\
&\leq 2p_A(a)\left(\sum_{x\in \mathcal X}|g(x)|\right),\label{eq:alpha1}
\end{align}
where \eqref{eq:sim_proof_prob} follows from the property of the random mapping $F$ that $\Pr[F(x)=a]=p_A(a)$. Therefore, Equation~\eqref{eq:sim_claim2} holds for $\alpha=1$.

For $\alpha=2$ we have
\begin{align}
\sum_{f\in \mathcal F} &\left( p_F(f)\left |\sum_{x\in \mathcal X}g(x)\bigg( \mathbbm{1}(f(x)=a)-p_A(a) \bigg)\right|^2\right) \notag \\
&=\sum_{f\in \mathcal F} \left( p_F(f)\sum_{x,x'\in \mathcal X}g(x)g(x')\bigg( \mathbbm{1}(f(x)=a)-p_A(a) \bigg)\bigg( \mathbbm{1}(f(x')=a)-p_A(a) \bigg)\right)\notag\\
&=\sum_{f\in \mathcal F} \Bigg( p_F(f)\Bigg( \sum_{x\in \mathcal X}g(x)^2\bigg( \mathbbm{1}(f(x)=a)-p_A(a) \bigg)^2 + \notag\\
&\quad\quad\quad\sum_{x,x'\in \mathcal X, x\neq x'}g(x)g(x')\bigg( \mathbbm{1}(f(x)=a)-p_A(a) \bigg)\bigg( \mathbbm{1}(f(x')=a)-p_A(a) \bigg) \Bigg)\Bigg)\notag\\
&=\sum_{x\in \mathcal X}\left(g(x)^2\sum_{f\in \mathcal F}\left(p_F(f)\bigg( \mathbbm{1}(f(x)=a)-p_A(a) \bigg)^2\right)\right) \label{eq:sim_proof_indep}\\
&=\sum_{x\in \mathcal X}\left(g(x)^2\bigg(p_A(a)(1-p_A(a))^2+(1-p_A(a))p_A(a)^2\bigg)\right)\label{eq:sim_proof_simplify}\\
&=p_A(a)(1-p_A(a))\sum_{x\in \mathcal X}g(x)^2\notag\\
&\leq p_A(a)\sum_{x\in \mathcal X}g(x)^2,\notag
\end{align}
where \eqref{eq:sim_proof_indep} follows from the fact that for distinct $x$ and $x'$, $F(x)$ is independent of $F(x')$, and $\sum_{f\in \mathcal F}P_F(f)\mathbbm{1}(f(x)=a) = \Pr[F(x)=a] = P_A(a)$. Equation \eqref{eq:sim_proof_simplify} is implied by $\Pr[F(x)=a]=p_A(a)$.

\setcounter{equation}{0}
\section{Proof of Proposition~\ref{pro:simulation}}\label{sec:proof_simulation2}
$F$ is fully described by its elements $\{F(x)\}_{x\in \mathcal X}$, and hence, $D_{TV}$ is a deterministic function of $\{F(x)\}_{x\in \mathcal X}$, \emph{i.e.,}
$$D_{TV}=g(\{F(x)\}_{x\in \mathcal X}),$$
where $g:\mathcal A^{|\mathcal X|}\to \mathbbm{R}$ is a deterministic function defined as follows:
$$g(\{f(x)\}_{x\in \mathcal X})=\|p_{f(X)Y}-p_Ap_Y\|_{TV}.$$
Since $D_{TV}$ is a function of independent random variables, we utilize the McDiarmid's inequality (\cite{mcdiarmid}).

Let $f,\tilde f\in \mathcal F$ be two arbitrary mappings with equal assignments for all elements of $\mathcal X$ except for some element $x_0$, \emph{i.e.,}
$$f(x)=\tilde f(x), \forall x\in \mathcal X/\{x_0\},\quad f(x_0)\neq \tilde f(x_0).$$
Then, we have:
\begin{align*}
g(&\{\tilde f(x)\}_{x\in \mathcal X})=\frac 12\sum_{y\in \mathcal Y}\left(p(y)\sum_{a\in \mathcal A}\left|\sum_{x\in \mathcal X}p(x|y)\mathbbm{1}[\tilde f(x)=a]-p(a)\right|\right)\\
&=\frac 12\sum_{y\in \mathcal Y}\left(p(y)\sum_{a\in \mathcal A}\left|\sum_{x\in \mathcal X} p(x|y)\mathbbm{1}[f(x)=a]-p(a)+p(x_0|y)(\mathbbm{1}[\tilde f(x_0)=a]-\mathbbm{1}[f(x_0)=a])\right|\right)\\
&\leq g(\{f(x)\}_{x\in \mathcal X})+ \frac 12\sum_{y\in \mathcal Y}\left(p(y)\sum_{a\in \mathcal A}\left|p(x_0|y)(\mathbbm{1}[\tilde f(x_0)=a]-\mathbbm{1}[f(x_0)=a])\right|\right)\\
&= g(\{f(x)\}_{x\in \mathcal X})+ p(x_0).
\end{align*}
Therefore, we have
$$|g(\{\tilde f(x)\}_{x\in \mathcal X})-g(\{f(x)\}_{x\in \mathcal X})|\leq p_X(x_0).$$
Furthermore, recall that $\{F(x)\}_{x\in \mathcal X}$ are independent random variables. Hence, McDiarmid's inequality (\cite{mcdiarmid}) implies:
$$\Pr[|g(\{F(x)\}_{x\in \mathcal X})-\E[g(\{F(x)\}_{x\in \mathcal X})]|\geq t]\leq 2e^{-\frac{2t^2}{\sum_{x\in\mathcal X}p(x)^2}}=2e^{-2t^22^{H_2(X)}}.$$

\setcounter{equation}{0}
\section{Proof of Theorem~\ref{theorem:autonomous}}\label{sec:proof_autonomous}
In \ref{sec:proof_ach} we show that provided the conditions of Theorem \ref{theorem:autonomous}, $(\epsilon_A,\epsilon_B)$-Nash equilibrium is achievable by autonomous strategies (achievability proof), and in \ref{sec:proof_conv}, we show that if the autonomous strategy profile $(\sigma^n,\tau^n)$ forms an $(\epsilon_A,\epsilon_B)$-Nash equilibrium, then, $\epsilon_A$ and $\epsilon_B$ satisfy the conditions of Theorem~\ref{theorem:autonomous} (converse proof).
\subsection{Achievability proof}\label{sec:proof_ach}
Let $A$, $B$ and $Q$ be the random variables in the statement of Theorem~\ref{theorem:autonomous} for which the entropy constraints hold strictly, \emph{i.e.,} $H(A|Q)<H(X)$, and $H(B|Q)<H(Y)$. Take $n$ of the form $n=NL$, and divide the total $n$ stages into $L$ blocks of $N$ stages. We generate the action sequence of each block (except for the first block) as a function of the random source observed during the previous block. In all stages of the first block, fixed actions $a\in \mathcal A$ and $b\in \mathcal B$ are played by Alice and Bob, respectively. Furthermore, excluding the first block, each block is further divided into four subblocks each of which include the following set of stages:
\begin{align}
&\mathcal{I}_0=\{1,2,\dots,\lceil p_Q(0)N\rceil \},\notag\\
&\mathcal{I}_1=\{\lceil p_Q(0)N\rceil+1,\lceil p_Q(0)N\rceil+2,\dots,\lceil p_Q(0)N\rceil+\lceil p_Q(1)N\rceil \},\notag\\
&\mathcal{I}_2=\{\lceil p_Q(0)N\rceil+\lceil p_Q(1)N\rceil+1,\dots,\lceil p_Q(0)N\rceil+\lceil p_Q(1)N\rceil+\lceil p_Q(2)N\rceil \},\notag\\
&\mathcal{I}_3=\{\lceil p_Q(0)N\rceil+\lceil p_Q(1)N\rceil+\lceil p_Q(2)N\rceil+1,\dots,N \}.\label{eq:subblock}
\end{align}
We generate strategies of Alice and Bob in such a way that they use the randomness source observed in last block to generate the actions of current block. We would like the generated actions of Alice and Bob to be almost \emph{i.i.d.\@} according to respective distributions $p_{A|Q=q}$ and $p_{B|Q=q}$ during each subblock $\mathcal I_q$ for all $q\in\{0,1,2,3\}$. Moreover, the action played in each stage should be also independent of the other player's observations up to that stage. As $H(A|Q)< H(X)$ and $H(B|Q)< H(Y)$, intuitively, each player could generate his/her actions with the intended pmf as a function of his/her corresponding randomness source observed during the previous block. Then, $\epsilon ^{(A)} (A,B|Q)\leq \epsilon_A$ and $\epsilon ^{(B)} (A,B|Q)\leq \epsilon_B$ would imply that in limit, the constructed strategies form an $(\epsilon_A,\epsilon_B)$-Nash equilibrium. We will now present the above sketch of proof more precisely.

For arbitrary $i\in\{1,2,\dots,L\}$, let $A_i^N=(A_{i,1},\dots,A_{i,N})$ and $B_i^N=(B_{i,1},\dots,B_{i,N})$ denote the actions of Alice and Bob, respectively, and let $X_i^N=(X_{i,1},\dots,X_{i,N})$ and $Y_i^N=(Y_{i,1},\dots,Y_{i,N})$ be the randomness sources observed by Alice and Bob, respectively, during block number $i$. Let us construct the strategies $\sigma^n$ for Alice and $\tau^n$ for Bob as follows: Alice and Bob choose their actions in each block $i\geq 2$ as a deterministic function of the corresponding sequence of random sources observed during the previous block, \emph{i.e.,} block number $i-1$. Particularly, there exist a sequence of deterministic mappings $\{\varphi_i\}_{i=2}^L$ and $\{\psi_i\}_{i=2}^L$ such that for all $i\geq 2$ we have $A_i^N=\varphi_i(X_{i-1}^N)$, and $B_i^N=\psi_i(Y_{i-1}^N)$. In all stages of the first block, both Alice and Bob choose an arbitrary fixed action. Next, we fulfill the specification of the strategies $\sigma^n$ and $\tau^n$ by specifying the functions $\{\varphi_i\}_{i=2}^L$ and $\{\psi_i\}_{i=2}^L$.

Let $q_{A^N}$ and $q_{B^N}$ be ideal distributions defined as follows:
\begin{equation}\label{eq:def_q}
q_{A^N}(a^N)= \prod_{q=0}^3\prod_{t\in \mathcal{I}_q}p_{A|Q}(a_{t}|q),\quad q_{B^N}(b^N)= \prod_{q=0}^3\prod_{t\in \mathcal{I}_q}p_{B|Q}(b_{t}|q).
\end{equation}
For arbitrary $i\geq 2$, let $\varphi_i$ be the mapping of Lemma~\ref{lemma:simulation} that simulates $q_{A^N}$ from $X_{i-1}^N$, and let $\psi_i$ be the mapping of Lemma~\ref{lemma:simulation} that simulates $q_{B^N}$ from $Y_{i-1}^N$. Thus, for all $i\geq 2$ and $1\leq \alpha\leq 2$, we have
\begin{align}
\left\|p_{A_i^N}-q_{A^N}\right\|_{TV} \leq 2^{-(1-\frac{1}{\alpha})\left(H_{\alpha}(X_{i-1}^N)-H_{\frac{1}{\alpha}}(q_{A^N})+2\right)}, \label{eq:tv0a} \\
\left\|p_{B_i^N}-q_{B^N}\right\|_{TV} \leq 2^{-(1-\frac{1}{\alpha})\left(H_{\alpha}(Y_{i-1}^N)-H_{\frac{1}{\alpha}}(q_{B^N})+2\right)}. \label{eq:tv0b}
\end{align}
Note that $H_{\alpha}(X_{i-1}^N)=NH_{\alpha}(X)$, and
\begin{align*}
H_{\frac{1}{\alpha}}(q_{A^N})=\sum_{q=0}^3 |\mathcal I_q|H_{\frac{1}{\alpha}}(p_{A|Q=q})&\leq 4\log |\mathcal A|+N\sum_{q=0}^3 p_{Q}(q) H_{\frac{1}{\alpha}}(p_{A|Q=q})\\
&\leq N\left (2+\sum_{q=0}^3 p_{Q}(q) H_{\frac{1}{\alpha}}(p_{A|Q=q})\right),
\end{align*}
where the first inequality follows from $|\mathcal I_q|\leq p_Q(q)N+1$ and $H_{\frac{1}{\alpha}}(p_{A|Q=q})\leq \log |\mathcal A|$; the second inequality holds for sufficiently large $N$. Therefore, Equation~\eqref{eq:tv0a} implies
\begin{equation}\label{eq:tv_temp}
\left\|p_{A_i^N}-q_{A^N}\right\|_{TV} \leq 2^{-N(1-\frac{1}{\alpha})\left(H_{\alpha}(X)-\sum_{q=0}^3 p_{Q}(q) H_{\frac{1}{\alpha}}(p_{A|Q=q})\right)}.
\end{equation}
Note that $\lim_{\alpha\to 1} \left\{H_{\alpha}(X)-\sum_{q=0}^3 p_{Q}(q) H_{\frac{1}{\alpha}}(p_{A|Q=q})\right\} = H(X)-H(A|Q)>0$; thus, there exits a $\alpha>1$ such that $H_{\alpha}(X)-\sum_{q=0}^3 p_{Q}(q) H_{\frac{1}{\alpha}}(p_{A|Q=q})>0$. Therefore, Equation~\eqref{eq:tv_temp} implies that for arbitrary $\delta>0$, one can choose $N$ large enough so that
\begin{equation} \label{eq:tv1a}
\left|\left| p_{A_i^N}-q_{A^N}\right|\right|_{TV} \leq \delta.
\end{equation}
A similar argument concludes that for sufficiently large $N$ we have
\begin{equation} \label{eq:tv1b}
\left|\left| p_{B_i^N}-q_{B^N}\right|\right|_{TV} \leq \delta.
\end{equation}

Note that $X_{i-1}^N$ is independent of $Y_{i-1}^N$, thus, $A_i^N$ is independent of $B_i^N$, \emph{i.e.,} $p_{A_i^NB_i^N}=p_{A_i^N}p_{B_i^N}$. This fact along with Equations~\eqref{eq:tv1a}, \eqref{eq:tv1b} and \eqref{eq:def_q} implies
\begin{equation}\label{eq:tv2}
\left|\left| p_{A_i^NB_i^N}(a^N,b^N)-\prod_{q=0}^3\prod_{t\in \mathcal{I}_q}p_{A|Q}(a_{t}|q)p_{B|Q}(b_{t}|q)\right|\right|_{TV} \leq 2\delta,
\end{equation}
where we utilized the third property of total variation in Lemma~\ref{lemma:tv_p}. Note that $p_{A_i^NB_i^N}(a^N,b^N)$ is the actual distribution of actions in block number $i$, while $\prod_{q=0}^3\prod_{t\in \mathcal{I}_q}p_{A|Q}(a_{t}|q)p_{B|Q}(b_{t}|q)$ is the ideal one. Next, for arbitrary $\delta'>0$, we have
\begin{align}
\lambda(\sigma^n,\tau^n)&=\frac{1}{n}\sum_{t=1}^n \E_{\sigma^n,\tau^n}[u_{A_tB_t}] \notag \\
&\geq \frac{1}{NL}\left( -\mathsf{M}N+(L-1)\sum_{q=0}^3 |\mathcal{I}_q|\E[u_{AB}|Q=q]-4(L-1)N\mathsf{M}\delta\right) \notag \\
& \geq \E[u_{AB}]-\delta'-4\mathsf{M}\delta, \label{eq:payoff1}
\end{align}
where the first inequality follows from the following two facts:
\begin{enumerate}
\item The expected payoff of the first block is bounded below by $-\mathsf MN$, where $\mathsf M=\max_{(a,b)\in \mathcal A\times \mathcal B}|u_{ab}|$.
\item Inequality~\ref{eq:tv2} implies that excluding the first block, the expected payoff of each block is in at most $4\mathsf MN \delta$ distance of the expected payoff induced by the ideal distribution of actions.
\end{enumerate}
The second inequality holds for sufficiently large $L$ and $N$, because $(L-1)|\mathcal I_q|/NL$ tends to $p_Q(q)$, and $\mathsf M/L$ tends to zero as $N$ and $L$ tend to infinity.

Next, we prove that the strategies $\sigma^n$ and $\tau^n$ constructed above form the desired approximate Nash equilibrium. Consider an arbitrary strategy $\hat{\sigma}^n$ (not necessarily an autonomous strategy) for Alice, and let Alice and Bob play according to strategy profile $(\hat{\sigma}^n,\tau^n)$. In this case, let $\hat{A}^N_i$ and $\hat{B}^N_i$ denote the sequence of actions of Alice and Bob in block number $i\geq 1$; moreover, let $\hat{X}_{i}^N$ and $\hat{Y}_i^N$ denote the sequence of randomness sources observed during block number $i\geq 1$. Observe that $\tau^n$ is an autonomous strategy, thus, changing the strategy of Alice from $\sigma^n$ to $\hat{\sigma}^n$ has not any impact on the actions of Bob; hence, the sequence of actions of Bob $(\hat{B}^N_i)$ still satisfies the property of Equation~\eqref{eq:tv1b}, \emph{i.e.,}
\begin{equation}\label{eq:tv3}
\left\| p_{\hat{B}^N_i}(b^N)-\prod_{q=0}^3\prod_{t\in \mathcal{I}_q}p_{B|Q}(b_{t}|q)\right\|_{TV} \leq \delta,\forall i=2,\dots,L.
\end{equation}

Note that at $t$-th stage of block number $i$, Alice finds information about $\hat{Y}_{i-1}^N$ just through $\hat{B}_{i}^{t-1}$; thus, $\hat{A}_{i,t}$ is independent of $\hat{Y}_{i-1}^N$ given $\hat{B}_{i}^{t-1}$. On the other hand, $(\hat{B}_{i,t},\hat{B}_{i,t+1},\dots,\hat{B}_{i,N})$ is a deterministic function of $\hat{Y}_{i-1}^N$. Therefore, $\hat{A}_{i,t}$ is also independent of $(\hat{B}_{i,t},\hat{B}_{i,t+1},\dots,\hat{B}_{i,N})$ given $\hat{B}_{i}^{t-1}$. Hence,

\begin{equation}\label{eq:markov}
p_{\hat{A}^N_i|\hat{B}^N_i}(a^N|b^N)=\prod_{t=1}^N p_{\hat{A}_{i,t}|\hat{A}^{t-1}_i\hat{B}^{t-1}}(a_{t}|a^{t-1},b^{t-1}).
\end{equation}

Equations~\eqref{eq:tv3} and \eqref{eq:markov} along with the first property of total variation in Lemma~\ref{lemma:tv_p} imply that for all $i\geq 2$, we have
\begin{equation}\label{eq:tv4}
\left|\left| p_{\hat{A}^N_i\hat{B}^N_i}(a^N,b^N)-\prod_{q=0}^3\prod_{t\in \mathcal{I}_q}p_{B|Q}(b_{t}|q)p_{\hat{A}_{i,t}|\hat{A}^{t-1}_i\hat{B}^{t-1}_i}(a_{t}|a^{t-1},b^{t-1}) \right|\right|_{TV} \leq \delta.
\end{equation}
Note that the ideal distribution $\prod_{q=0}^3\prod_{t\in \mathcal{I}_q}p_{B|Q}(b_{t}|q)p_{\hat{A}_{i,t}|\hat{A}^{t-1}_i\hat{B}^{t-1}_i}(a_{t}|a^{t-1},b^{t-1})$ guarantees that the expected payoff of block number $i\geq 2$ is no more than
$$\sum_{q=0}^3 |\mathcal{I}_q|\max_{a\in \mathcal{A}}\E[u_{aB}|Q=q].$$
On the other hand, Equation~\eqref{eq:tv4} implies that the actual expected payoff of arbitrary block number $i\geq 2$ is in $2\mathsf MN \delta$ distance of the ideal one. Thus, the actual expected payoff of block number $i\geq 2$ is no more than
$$\sum_{q=0}^3 |\mathcal{I}_q|\max_{a\in \mathcal{A}}\E[u_{aB}|Q=q]+2\mathsf M N \delta.$$
Furthermore, the expected payoff of the first block is bounded above by $N\mathsf M$; thus,
\begin{align}
\lambda(\hat{\sigma}^n,\tau^n)&=\frac{1}{n}\sum_{t=1}^n \E_{\hat{\sigma}^n,\tau^n}[u_{\hat{A}_t\hat{B}_t}] \notag \\
&\leq \frac{1}{NL}\left( \mathsf{M}N+(L-1)\sum_{q=0}^3 |\mathcal{I}_q|\max_{a\in \mathcal{A}}\E[u_{aB}|Q=q]+2(L-1)N\mathsf{M}\delta\right) \notag \\
& \leq \delta'+2\mathsf{M}\delta+\sum_{q=0}^3 p_Q(q)\left[ \max_{a\in \mathcal{A}} \E[u_{aB}|Q]\right], \label{eq:payoff2}
\end{align}
where the second inequality holds for sufficiently large $L$ and $N$.

Equations \ref{eq:payoff1} and \ref{eq:payoff2} conclude
\begin{align}
\lambda(\hat{\sigma}^n,\tau^n) - \lambda(\sigma^n,\tau^n) &\leq \sum_{q=0}^3 p_Q(q) \left[ \max_{a\in \mathcal{A}} \E[u_{aB}|Q]\right]-\E[u_{AB}]+2\delta'+6\mathsf{M}\delta \notag \\
&=g_A(A,B|Q)+2\delta'+6\mathsf{M}\delta \notag \\
& \leq \epsilon_A + 2\delta'+6\mathsf{M}\delta \label{eq:improve_A}
\end{align}
where the second inequality follows from the assumption in the statement of the theorem that $g_A(A,B|Q)\leq \epsilon_A$. By a similar argument as above, we can show that for arbitrary strategy $\hat{\tau}^n$ for Bob we have
\begin{equation}\label{eq:improve_B}
\lambda(\sigma^n,\tau^n)-\lambda(\sigma^n,\hat{\tau}^n)\leq \epsilon_B + 2\delta'+6\mathsf{M}\delta.
\end{equation}
Inequalities \eqref{eq:improve_A} and \eqref{eq:improve_B} imply that the strategy profile $(\sigma^n,\tau^n)$ forms a $(\epsilon_A+2\delta'+6\mathsf{M}\delta,\epsilon_B+2\delta'+6\mathsf{M}\delta)$-Nash equilibrium. But one can choose $\delta$ and $\delta'$ small enough to make $2\delta'+6\mathsf{M}\delta$ as small as desired; thus, $(\epsilon_A,\epsilon_B)$-Nash equilibrium is achievable.

\subsection{Converse proof}\label{sec:proof_conv}
Let $(\sigma^n,\tau^n)$ be an autonomous strategy profile generating an $(\epsilon_A,\epsilon_B)$-Nash equilibrium in the $n$ stage repeated game, and let $A^n$ and $B^n$ be the sequence of actions of Alice and Bob. Let $T$ be a random variable chosen from $\{1,2,\dots,n\}$ uniformly and independent of $(X^n,Y^n,A^n,B^n)$. Let us define
$$\tilde{A}=A_T,\quad \tilde{B}=B_T, \quad Q=(T,A^{T-1},B^{T-1}).$$
We show that the random variables $Q, \tilde{A}, \tilde{B}$ along with $\epsilon_A$ and $\epsilon_B$ satisfy the conditions of Theorem~\ref{theorem:autonomous}.

\textbf{The Markov conditions:} In this part, our goal is to show that $\tilde{A}$ is independent of $\tilde{B}$ given $Q$,\emph{i.e.,} $p_{\tilde{A}\tilde{B}Q}=p_Q p_{\tilde{A}|Q}p_{\tilde{B}|Q}$. To do this, it suffices to prove that for all $t\geq 1$, $A_t$ is independent of $B_t$ given $(A^{t-1},B^{t-1})$. Note that the strategies $\sigma^n$ and $\tau^n$ are autonomous, hence, $A^n$ is a deterministic function of $X^n$, and $B^n$ is a deterministic function of $Y^n$; thus, the fact that $X^n$ is independent of $Y^n$ implies that $A^n$ is independent of $B^n$. Therefore, $A_t$ is independent of $B_t$ given $(A^{t-1},B^{t-1})$.

\textbf{Entropy conditions:} We show that $H(\tilde{A}|Q)\leq H(X)$:
\begin{align}
H(\tilde{A}|Q)&=H(A_T|T,A^{T-1},B^{T-1})\notag\\
&=\frac{1}{n}\sum_{t=1}^n H(A_t|A^{t-1},B^{t-1}) \label{eq:entropy1}\\
&=\frac{1}{n}\sum_{t=1}^n H(A_t|A^{t-1}) \label{eq:entropy2}\\
&=\frac{1}{n}H(A^n) \notag\\
&\leq \frac{1}{n}H(X^n)\label{eq:entropy4}\\
&= H(X)\notag
\end{align}
where \eqref{eq:entropy1} follows from the independence of $T$ from $(A^n,B^n)$. Recall that $X^n$ is independent of $Y^n$, thus, $A^n$ --a deterministic function of $X^n$-- is independent of $B^n$ --a deterministic function of $Y^n$. Hence, Equation~\eqref{eq:entropy2} is correct. Equation~\eqref{eq:entropy4} is implied by the fact that $A^n$ is a deterministic function of $X^n$. A similar argument justifies $H(\tilde{B}|Q)\leq H(Y)$.

\textbf{Equilibrium conditions:} In this part we show that $g_A(\tilde{A},\tilde{B}|Q) \leq \epsilon_A$. To do this, we consider a new game in which Alice and Bob play according to strategy profile $(\hat{\sigma}^n,\tau^n)$, where $\hat{\sigma}^n$ will now be constructed. In the new game, let $\hat{A}^n$ and $\hat{B}^n$ denote the respective actions of Alice and Bob; furthermore, let $\hat{X}^n$ and $\hat{Y}^n$ denote the respective sources of randomness of Alice and Bob, and $\hat{\mathsf H}^{t}_1$ denote the history of observations of Alice until stage $t$. Note that $\tau^n$ is the same strategy as considered in the beginning of the converse proof, whereas $\hat{\sigma}^n=(\hat{\sigma}_1,\dots,\hat{\sigma}_n)$ is generated as follows: given $\hat{h}_1^t$, an arbitrary history of observations of Alice until stage $t$, $\hat{\sigma}_t(\hat{h}_1^t)$ is the best choice of Alice that maximizes the expected payoff at stage $t$, \emph{i.e.,}
$$ \hat{\sigma}_t(\hat{h}_1^t) \in \underset{a\in \mathcal{A}}{\arg \max} \textrm{  }\E_{\tau^n} \left[u_{a\hat{B}_t}|\hat{\mathsf{H}}_1^t=\hat{h}_1^t\right],$$
where $\E_{\tau^n}$ denotes the expectation with respect to the probability distribution induced by $\tau^n$ and $p_{Y}$. We have
\begin{align}
\lambda(\hat{\sigma}^n,\tau^n) &= \sum_{t=1}^n \sum_{\hat{h}_1^t}\frac{1}{n} p_{\hat{\mathsf{H}}_1^t}(\hat{h}_1^t)\max_{a\in\mathcal{A}} \E_{\tau^n}[u_{a\hat{B}_t}|\hat{\mathsf{H}}_1^t=\hat{h}_1^t] \notag \\
&=  \sum_{t=1}^n \sum_{\hat{x}^t,\hat{a}^{t-1},\hat{b}^{t-1}}\frac{1}{n} p_{\hat{X}^t\hat{A}^{t-1}\hat{B}^{t-1}}(\hat{x}^t,\hat{a}^{t-1},\hat{b}^{t-1})\max_{a\in\mathcal{A}} \E_{\tau^n}[u_{a\hat{B}_t}|\hat{x}^t,\hat{a}^{t-1},\hat{b}^{t-1}] \notag \\
&=  \sum_{t=1}^n \sum_{\hat{x}^t,\hat{a}^{t-1},\hat{b}^{t-1}}\frac{1}{n} p_{\hat{X}^t\hat{A}^{t-1}\hat{B}^{t-1}}(\hat{x}^t,\hat{a}^{t-1},\hat{b}^{t-1})\max_{a\in\mathcal{A}} \E_{\tau^n}[u_{a\hat{B}_t}|\hat{b}^{t-1}]\label{eq:equilibria1}\\
&=  \sum_{t=1}^n \sum_{\hat{b}^{t-1}}\frac{1}{n} p_{\hat{B}^{t-1}}(\hat{b}^{t-1})\max_{a\in\mathcal{A}} \E_{\tau^n}[u_{a\hat{B}_t}|\hat{b}^{t-1}] \notag \\
&=  \sum_{t=1}^n \sum_{b^{t-1}}\frac{1}{n} p_{B^{t-1}}(b^{t-1})\max_{a\in\mathcal{A}} \E_{\tau^n}[u_{aB_t}|b^{t-1}] \label{eq:equilibria2}\\
&=  \sum_{t=1}^n \sum_{a^{t-1},b^{t-1}}\frac{1}{n} p_{A^{t-1}B^{t-1}}(a^{t-1},b^{t-1})\max_{a\in\mathcal{A}} \E_{\tau^n}[u_{aB_t}|a^{t-1},b^{t-1}] \label{eq:equilibria3}\\
&=\sum_{q}p_Q(q) \left[ \max_{a\in \mathcal{A}} \E[u_{a\tilde B}|Q]\right]\label{eq:equilibria4},
\end{align}
where Equation~\eqref{eq:equilibria1} results from the following two facts:
\begin{enumerate}
\item $\hat{A}^{t-1}$ is a deterministic function of $(\hat{X}^{t},\hat{B}^{t-1})$, thus, $\hat{B}_t$ is independent of $\hat{A}^{t-1}$ given $(\hat{X}^{t},\hat{B}^{t-1})$.
\item $\tau^n$ is an autonomous strategy, thus, $\hat{B}^{t}$ is a deterministic function of $\hat{Y}^t$. On the other hand, $\hat{Y}^t$ is independent of $\hat{X}^t$. Therefore, $\hat{B}^{t}$ is independent of $\hat{X}^t$.
\end{enumerate}
Note again that $\tau^n$ is an autonomous strategy, thus, the probability distribution of the actions of Bob is not related to the strategy of Alice; hence, $p_{\hat{B}^n}=p_{B^n}$ (recall that $B^n$ is the sequence of actions of Bob in the original game in which strategy profile $(\sigma^n,\tau^n)$ is played); as a result, Equation~\eqref{eq:equilibria2} holds. Since $\sigma^n$ and $\tau^n$ are autonomous strategies, $A^n$ is a deterministic function of $X^n$, and $B^n$ is a deterministic function of $Y^n$. This fact along with the independence of $X^n$ from $Y^n$ implies that $A^n$ is independent of $B^n$, thus Equation~\eqref{eq:equilibria3} follows.

Furthermore, the expected average payoff induced by $(\sigma^n,\tau^n)$ equals
\begin{equation}\label{eq:equilibria5}
\lambda(\sigma^n,\tau^n) = \sum_{t=1}^n \frac{1}{n} \E_{\tau^n}[u_{A_tB_t}] = \E[u_{\tilde{A}\tilde{B}}] = \sum_{q}p_Q(q)\E[u_{\tilde{A}\tilde{B}}|Q].
\end{equation}

Equations~\eqref{eq:equilibria4} and \eqref{eq:equilibria5} imply
\begin{align*}
\lambda(\hat{\sigma}^n,\tau^n)-\lambda(\sigma^n,\tau^n) = \sum_{q}p_Q(q) \left[ \max_{a\in \mathcal{A}} \E[u_{aB}|Q]-\E[u_{\tilde{A}\tilde{B}}|Q]\right] = g_A(A,B|Q).
\end{align*}
The above equation along with the fact that $(\sigma^n,\tau^n)$ is an $(\epsilon_A,\epsilon_B)$-Nash equilibrium implies that $g_A(A,B|Q)\leq \epsilon_A$. Using a similar argument as above, one can show that $g_B(A,B|Q) \leq \epsilon_B$.

\textbf{Cardinality bound:} The identified random variables $(Q,\tilde{A},\tilde{B})$ satisfy the constraints of the theorem, except the cardinality bound on $Q$. Cardinality of $Q$ can be reduced using standard arguments such as the support lemma of \cite[Appendix C]{elgamal}, or the Fenchel-Bunt extension to the Caratheodory's theorem. We modify $p_{Q\tilde A\tilde B}$ and generate a new distribution $p'_{Q\tilde A\tilde B}$ so that it also satisfies the cardinality bound. Let $p'_{\tilde A\tilde B|Q}=p_{\tilde A\tilde B|Q}$; this guarantees that under $p'$, given $Q$, $\tilde A$ is independent of $\tilde B$. Next, we complete the definition of $p'_{Q\tilde A\tilde B}$ by specifying the marginal distribution $p'_{Q}$. We can perceive $p'_{Q}$ as a real vector $[p'_{Q}(q),q\in \mathcal Q]$ satisfying the following linear constraints:
\begin{align}
&p'_Q(q)\geq 0, q\in \mathcal Q\label{eq:cardinality_simplex1},\\
&\sum_{q\in \mathcal Q} p'_Q(q)=1\label{eq:cardinality_simplex2},
\end{align}
where $\mathcal Q$ is the sample space of the random variable $Q$.

Let $H'(\tilde A|Q)$ denote the entropy of $\tilde A$ given $Q$, under the distribution $p'_{\tilde A\tilde B Q}$. Similarly, the prime superscript in $H'(\tilde B|Q)$, $g'_A(\tilde A,\tilde B|Q)$ and $g'_B(\tilde A,\tilde B|Q)$ indicates that they are computed according to probability distribution $p'_{\tilde A\tilde B Q}$. We drop the superscript when the evaluation is done under the original probability distribution $p_{\tilde A\tilde B Q}$. Assume that $p'_{Q}$ also satisfies the following linear constraints:
\begin{align}
&H'(\tilde A|Q)=H(\tilde A|Q)\label{eq:cardinality_poly1},\\
&H'(\tilde B|Q)=H(\tilde B|Q)\label{eq:cardinality_poly2},\\
&g'_A(\tilde A,\tilde B|Q)=g_A(\tilde A,\tilde B|Q)\label{eq:cardinality_poly3}.
\end{align}
Let $\mathsf P$ denote the polytope of marginal distributions $p'_Q$ satisfying \eqref{eq:cardinality_simplex1}-\eqref{eq:cardinality_poly3}. Note that $\mathsf P$ is not empty since it contains $p_Q$. We choose $p'_Q$ to be an element of $\mathsf P$ that minimizes $g'_B(\tilde A,\tilde B|Q)$. This guarantees that $p'_{\tilde A\tilde B Q}$ inherits the following properties from $p_{\tilde A\tilde B Q}$:
\begin{align*}
&H'(\tilde A|Q) \leq H(X), \quad H'(\tilde B|Q)\leq H(Y), \\
&g'_A(\tilde A,\tilde B|Q)\leq\epsilon_A, \quad g'_B(\tilde A,\tilde B|Q) \leq \epsilon_B.
\end{align*}
We will now show that $p'_{\tilde A\tilde B Q}$ also satisfies the cardinality bound on the support of $Q$. Note that $g'_B(\tilde A,\tilde B|Q)$ is linear in $p'_Q$, hence, it's minimum occurs in a vertex of the polytope $\mathsf P$. polytope $\mathsf P$ lies in a $|\mathcal Q|$ dimensional space, and hence, each of it's vertices lies in at least $|\mathcal Q|$ out of the $|\mathcal Q|+4$ hyperplanes defining $\mathsf P$ (Equations~\eqref{eq:cardinality_simplex1}-\eqref{eq:cardinality_poly3}) . Therefore, $p'_Q$, which is a vertex of $\mathsf P$, lies in at least $|\mathcal Q|-4$ out of $|\mathcal Q|$ hyperplanes of the form \eqref{eq:cardinality_simplex1}. Hence, $p'_Q$ has at most 4 non-zero elements.

\appendix

\setcounter{equation}{0}
\renewcommand{\theequation}{\Alph{section}.\arabic{equation}}
%\numberwithin{equation}{section}
\renewcommand{\thesection}{Appendix \Alph{section}}

%\setcounter{equation}{0}
%\section{ff}\label{sec:appendix_ff}

%\section*{Acknowledgements}

%\section*{References}
\bibliographystyle{elsarticle-harv}
\bibliography{reference}


%\begin{thebibliography}{00}



%\end{thebibliography}

\end{document}